\documentclass[authoryear]{elsarticle}
\usepackage{latexsym}
\usepackage{amsfonts}
\usepackage{amsmath}
\usepackage{subfigure}
\usepackage{graphicx}
\usepackage{multirow}
\usepackage{lscape}
\usepackage{longtable}
\usepackage{bm}
\usepackage{color}
\usepackage{rotating}
\usepackage{verbatim}
\usepackage{natbib}
\bibpunct{(}{)}{,}{a}{}{,}

\newcommand{\MR}[1]{\hline \multicolumn{15}{l}{#1}\\}

\newcommand{\fx}{{\bf x}}

\newcommand{\ftheta}{\mbox{\boldmath$\theta$}}
\def\bma{{\boldsymbol a}}
\def\bmy{{\boldsymbol y}}
\def\bmu{{\boldsymbol u}}

\usepackage{setspace}
\begin{document}

\title{Modeling high dimensional time-varying dependence using D-vine SCAR models}

\author[ca]{Carlos Almeida}
\ead{carlos.almeida@uclouvain.be}
\address[ca]{Georges Lemaitre Centre for Earth and Climate Science,
Earth and Life Institute, Universit\'e catholique de Louvain, B-1348 Louvain-la-Neuve, Belgium.}

\author[cc]{Claudia Czado}
\ead{cczado@ma.tum.de}
\address[cc]{Department of Mathematical Statistics, Technische Univerit\"{a}t M\"{u}nchen, Germany.}

\author[hm]{Hans Manner\corref{cor}}
\ead{manner@statistik.uni-koeln.de}
\address[hm]{Department of Social and Economic Statistics, University
of Cologne, 50937 Cologne, Germany.}
\cortext[cor]{Corresponding author.}

\begin{frontmatter}

\begin{abstract}
We consider the problem of modeling the dependence among many time series. We build high dimensional time-varying copula models by
combining pair-copula constructions (PCC) with stochastic autoregressive copula (SCAR) models to capture
dependence that changes over time. We show how the estimation of this highly complex model can be broken down into the estimation of a sequence of
bivariate SCAR models, which can be achieved by using the method of simulated maximum likelihood. Further, by restricting the conditional dependence
parameter on higher cascades of the PCC to be constant, we can greatly reduce the number of parameters to be estimated without losing much flexibility.
We study the performance of our estimation method by a large scale Monte Carlo simulation. An application to a large dataset of stock returns of all
constituents of the Dax 30 illustrates the usefulness of the proposed model class.
\bigskip
\end{abstract}

\begin{keyword}
Stock return dependence, time-varying copula, D-vines, efficient importance sampling, sequential estimation

\emph{JEL Classification:} C15, C51, C58
\end{keyword}

\end{frontmatter}

\author{Carlos Almeida, Claudia Czado and Hans Manner}
\date{\today}

\section{Introduction}

The modeling of multivariate distributions is an important task for risk management and asset allocation problems. Since modeling the conditional mean of
financial assets is rather difficult, if not impossible, much research has focused on modeling conditional volatilities and dependencies. The
literature on multivariate GARCH \citep{BLR06} and stochastic volatility models \citep{HRS94,YM06} offers many approaches to extend univariate volatility
models to multivariate settings. However, usually the resulting multivariate model makes the assumption of (conditional) multivariate normality.
Multivariate models based on copulas offer a popular alternative in that non-elliptical multivariate distributions can be constructed in a tractable and
flexible way. The advantage of using copulas to construct multivariate volatility models is that one is free with the choice of the marginal model,
i.e. the univariate volatility model, and that it is possible to capture possibly asymmetric dependencies in the tails of different pairs of the distributions. In
particular, lower tail dependence often needs to be accounted for when measuring financial risks. Among many others, \citet{P09} or \citet{CLV04}, and
references therein, give an overview of copula based models in financial applications.

Two major drawbacks of most of the early applications of copula based models are that most studies focus on bivariate copulas
only, limiting the potential for real world applications, and that the dependence parameter is assumed to be time-constant. This is
in contrast to the empirically observed time-varying correlations. Each of these issues individually has been addressed in the
literature in recent years. Larger dimensional copulas other than Gaussian or Student copulas have become available through the
introduction of hierarchical Archimedean copulas by \citet{ST10} and \citet{OOS09}, factor copula models by \citet{OP11}, or the class of pair copula constructions by
\citet{ACFB09}. In particular the latter class, also called vine copula constructions, has become extremely popular because of
its flexibility and because of the possibility of estimating the large number of parameters sequentially. Examples of financial
applications of vine copula models are, e.g., \citet{CHV08}, \citet{dissmann-etal}, or \citet{BrechmannCzado2011}.
Copulas with time-varying parameters have been introduced by \citet{P06a} to model changing exchange rate dependencies.
Since then a number of studies have proposed different ways to specify time-varying copulas. For example, \citet{DE04}
test for structural breaks in copula parameters, \citet{GHS09} use a sequence of breakpoint tests to identify intervals
of constant dependence, \citet{GT11} and \citet{stoeber2011c} use a regime-switching model for changing dependencies, \citet{HR10} treat the
copula parameter as a smooth function of time and estimate it by local maximum likelihood, whereas \citet{HM10} and
\citet{AC10} proposed a model where the copula parameter is a transformation of a latent Gaussian autoregressive
process of order one. An overview and comparison of time-varying copula models is given in \citet{MR10}.
Only very few papers allow for time-varying parameters in larger dimensions. \citet{CHV08} estimate a regime-switching
vine copula and \citet{HV08} allow the parameter of a vine copula to be driven by a variation of the DCC model by \citet{E02}.

The contribution of this paper is to extend the stochastic autoregressive copula (SCAR) model by \citet{HM10} and \citet{AC10} to practically relevant
dimensions using D-vines. We discuss how the proposed model can be estimated sequentially using simulated maximum likelihood estimation. We also address
how a number of conditional copulas can be restricted to be time-constant or independence copulas without restricting the flexibility of the model
too much. Furthermore, we perform a large scale Monte Carlo study investigating the behavior of the proposed estimator and find that it shows an acceptable
performance. In our empirical study we apply our model to the returns of 29 constituents of the German DAX 30 and find that the model performs quite
well.

The remainder of the paper is structured as follows. The next section introduces copulas in general, SCAR models, D-vines copulas and shows how they can be
combined to obtain the flexible class of D-vine SCAR models. Section \ref{Sec:Estimation} treats the estimation of the proposed model, Section
\ref{Sec:Simulations} presents the results of our simulation study, Section \ref{Sec:Application} contains the empirical application and Section
\ref{Sec:Conclusion} gives conclusions and outlines further research.

\section{D-vine based SCAR models}\label{Sec:DVineSCAR}
We are interested in modeling the joint (conditional) distribution of a d-dimensional time series $\bmy_t=(y_{1,t},...,y_{d,t})$ for $t=1,...,T$.
We assume that each variable $y_{i,t}$ for $i=1,...,d$ follows an ARMA(p,q)-GARCH(1,1) process, i.e.
\begin{equation}
\label{ARMA-GARCH}
y_{i,t} = \beta_{i,0} + \sum_{j=1}^p \beta_{i,j} y_{i,t-j} + \sum_{k=1}^q \delta_{i,k} \sigma_{i,t-k} \varepsilon_{i,t-k} + \sigma_{i,t} \varepsilon_{i,t}
\end{equation}
with
\[
\sigma^2_{i,t}=\alpha_{i,0} + \alpha_{i,1} \varepsilon^2_{i,t-1} + \gamma_i \sigma^2_{i,t}.
\]
The usual stationarity conditions are assumed to hold. Denote the joint distribution of the standardized innovations $\varepsilon_{i,t}$ by
$G(\varepsilon_{1,t},...,\varepsilon_{d,t})$ and let their marginal distributions be $F_i(\varepsilon_{1,t}),...,F_d(\varepsilon_{d,t})$, respectively.
Then by Sklar's theorem there exists a copula $C$ such that
\begin{equation}
\label{full-model}
G(\varepsilon_{1,t},...,\varepsilon_{d,t})=C(F_i(\varepsilon_{1,t}),...,F_d(\varepsilon_{d,t})).
\end{equation}

Since all the marginal behavior is captured by the marginal distribution, the
copula captures the complete contemporaneous dependence of the distribution. Let
$u_{i,t}=F_i(\varepsilon_{i,t})$ be the innovations transformed to $U(0,1)$
random variable and define $\bmu_t:=(u_{1,t},...,u_{d,t})$.

In the remainder of this section we first show how time-varying dependence can be incorporated in bivariate copula models. Then we discuss how these models can be
extended to arbitrary dimensions, for which we need the notion of vines.

\subsection{Bivariate SCAR Copula Models}
For now, consider the bivariate time series process $(u_{i,t}, u_{j,t})$ for $t=1,...,T$. We assume that its distribution is given by
\begin{equation}
\label{biscarden}
(u_{i,t}, u_{j,t}) \sim C(\cdot,\cdot;\theta^{ij}_t)
\end{equation}
with $\theta^{ij}_t \in \Theta$ the time-varying parameter of the copula C. In order to be able to compare copula parameters that have
different domains, the copula can equivalently be parameterized in terms of Kendall's $\tau \in (-1,1)$.
This follows from the fact that for all copulas we consider there exists a one-to-one relationship between the copula parameter and Kendall's $\tau$,
which we express by $\theta_t^{ij}=r(\tau_t^{ij})$. We assume that $\tau^{ij}_t$ is driven by the latent Gausian AR(1) process $\lambda_t^{ij}$ given by
\begin{equation}
\lambda^{ij}_t=\mu_{ij}+\phi_{ij} (\lambda^{ij}_{t-1}-\mu_{ij})+\sigma_{ij} z_{ij,t},
\end{equation}
where $z_{ij,t}$ are independent standard normal innovations. We further assume $|\phi_{ij}|<1$ for stationarity and $\sigma_{ij}>0$ for identification.
Due to
the fact that $\lambda^{ij}_t$ takes values  on the real line we apply the inverse Fisher transform to map it into $(-1,1)$, the domain of $\tau^{ij}_t$:
\begin{equation}\label{eq:Inv_Fisher}
\tau^{ij}_t=\frac{exp(2 \lambda^{ij}_t)-1}{{exp(2 \lambda^{ij}_t)+1}}=:
\psi(\lambda^{ij}_t).
\end{equation}

\subsection{D-vine Distributions and Copulas}

While copulas are recognized as a very powerful tool to construct multivariate
distributions, in the past only the class of bivariate copulas (e.g.
\citealp{J97}  and \citealp{Nelsen}) was flexible enough to accommodate asymmetric
and/or tail dependence without placing unrealistic restrictions on the dependence structure. Recently  pair copula constructions (PCC) are
found to be very useful to construct flexible multivariate copulas. Here a
multivariate copula is built up with bivariate copula terms modeling
unconditional and conditional dependencies. The first such construction was proposed in
\citet{Joe96}. It was subsequently significantly extended to more general settings in
\citet{BC02}, \citet{Bedford2} and  \citet{vinebook}. They called the resulting
distributions regular (R) vines and explored them for the case of Gaussian pair
copulas. The backbone is a graphical representation in form of a sequence
of linked trees identifying the indices which make up the multivariate copula. In
particular, they proved that the specification of the corresponding pair copula
densities make up a valid multivariate copula density. Further properties,
estimation, model selection methods
and their use in complex modeling situations can be found in
\citet{kuro:joe:2010}.

\citet{ACFB09} recognized the potential of this construction for statistical
inference and developed  a sequential estimation (SE) procedure, which can be used as
starting values for  maximum likelihood estimation (MLE). \citet{BC02}
identified two interesting subclasses of regular vines called D-vines and
canonical (C)-vines. In the case of D-vines the sequence  of vine trees consist
of pair trees, while for C-vines they are starlike with a central node. This
shows that C-vines are more useful for data situations where the importance of
the variables can be ordered. This is not the case for the application we will
present later; therefore we concentrate on D-vines. However, we would like to note
that multivariate SCAR models can also be constructed based on C-vines and more generally
on R-vines.

Notably, C- and D-vines can be introduced from first principles (e.g.
\citealp{cza:2010}). For this let $(X_1,...,X_d)$ be a
set of variables with joint distribution $F$ and density $f$,
respectively. Consider the recursive decomposition
\begin{eqnarray}
f(x_1, \ldots,x_d)& = & \prod_{k=2}^d f(x_k|x_1,\ldots,x_{k-1}) \times
f(x_1). \label{decomp}
\end{eqnarray}
Here $F(\cdot|\cdot)$ and later
$f(\cdot|\cdot)$ denote conditional cdf's and densities,
respectively. As a second ingredient we utilize
 Sklar's theorem for dimension $d=2$ to express
the conditional density of $X_1$ given $X_2=x_2$ as
\begin{equation}
\label{sklarcon}
f(x_1|x_2)  =  c_{1 2 }(F_1(x_1), F_2(x_2)) \times f_1(x_1),
\end{equation}
where $ c_{1 2 }$ denotes an arbitrary  bivariate copula density.
For distinct indices $i,j,i_1,\cdots,i_k$ with $ i < j $ and  $i_1<\cdots
<i_k$  we now introduce the abbreviation
\begin{equation}
\label{abbrev}
 c_{i,j|D}:=c_{i,j|D}(F(x_i|\fx_D), F(x_j|\fx_D)),
\end{equation}
where $D:=\{i_1,\cdots,i_k\}$ and $\fx_D:=(x_{i_1},
\ldots, x_{i_k})$.
Using (\ref{sklarcon}) for the conditional distribution of $(X_1,X_k)$ given
$X_2=x_2,\ldots X_{k-1}=x_{k-1}$ we can express $f(x_k|x_1,\cdots,x_{k-1})$ recursively as
\begin{eqnarray}
f(x_k|x_{1},\ldots,x_{k-1}) & = &  c_{1,k|2:k-1}
\times f(x_{k}|x_2,\ldots, x_{k-1})\nonumber\\
& = & [\prod_{s=1}^{k-2}c_{s,k|s+1:k-1}]
\times c_{(k-1),k} \times f_k(x_k) \label{condichte},
\end{eqnarray}
where $r:s:=(r,r+1,\ldots,s)$ for integers $r$ and $s$ with $r<s$.
Using (\ref{condichte}) in (\ref{decomp}) and $s=i, k=i+j$ it follows that
\begin{eqnarray}
f(x_1,\ldots,x_d)
& = &[ \prod_{j=1}^{d-1} \prod_{i=1}^{d-j}
c_{i,i+j|i+1:i+j-1}] \cdot[\prod_{k=1}^d
f_k(x_k)] \label{dvine}
\end{eqnarray}
If the marginal distribution of $X_k$ are uniform for all $k=1,\cdots,d$, then
we call the corresponding density in (\ref{dvine}) a D-vine copula density and
the corresponding distribution function a D-vine copula.

\begin{figure}
\begin{center}
\includegraphics[scale=.5]{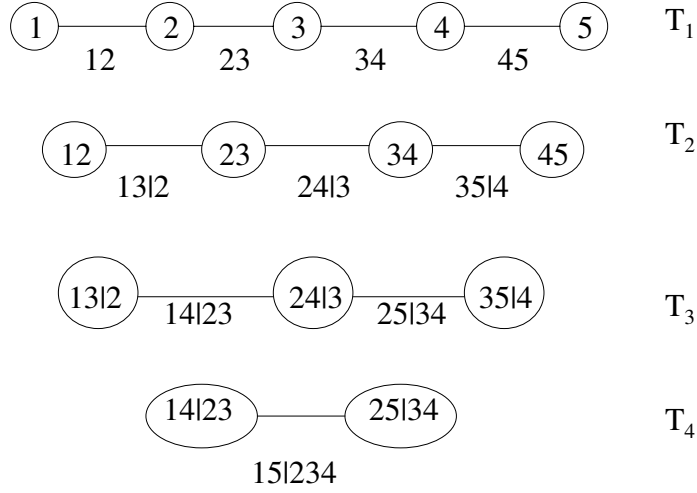}
\caption{A D-vine tree representation for $d=5$. }
\label{dvinetree}
\end{center}
\end{figure}

For illustration we consider a five dimensional D-vine, its density then
given by
\begin{eqnarray}
f(x_1,\cdots, x_5) & = & [\prod_{k=1}^5 f_k(x_k)] \cdot
c_{12} \cdot c_{23} \cdot c_{34} \nonumber \\
&\times &
  c_{45}
 \cdot c_{13|2}
 \cdot c_{24|3}  \cdot c_{35|4}
\cdot  c_{14|23} \cdot c_{25|34} \cdot c_{15|234},
\label{dvine5}
\end{eqnarray}
with the corresponding vine tree representation identifying the utilized
indices given in Figure \ref{dvinetree}. In particular the indices in
Tree $T_1$ indicate the unconditional pair copulas, while Trees $T_2$ to $T_4$
correspond to conditional pair copulas, where the set of conditioning variables
has size 1 to 3, respectively.

If $c_{i,i+j|i+1:i+j-1}$ models the dependence between the rv's
$F(X_i|\fx_{i+1:i+j-1})$ and \newline $F(X_{i+j}|\fx_{i+1:i+j-1})$ we implicitly
assume that the copula density $c_{i,i+j|i+1:i+j-1}(\cdot,\cdot)$
does not depend on the conditioning variables
$\fx_{i+1:i+j-1}$ other than through the arguments $F(X_i|\fx_{{i+1}:
{i+j-1}})$ and $F(X_{i+j}|\fx_{{i+1}:{i+j-1}})$. This is a common
assumption and \citet{haff-etal} call this a simplified vine. They showed that
this restriction is not severe by examining several examples.

In the D-vine representation given in (\ref{dvine}) we also need a fast
recursive way to compute conditional cdf's which enter as arguments. For this
\citet{Joe96} showed that for $v \in D$ and $D_{-v}:=D
\setminus v$
\begin{equation} \label{condDist}
F(x_j|\fx_D) =
\frac{\partial\,C_{j,v|D_{-v}}(F(x_j|\fx_{D_{-v}}),
F(x_v|\fx_{D_{-v}}))}{\partial F(x_v|\fx_{D_{-v}})}.
\end{equation}
For the special case of $D=\{v\}$ it follows that
\[
F(x_j|x_v) =
\frac{\partial\,C_{j,v}(F(x_j),
F(x_v))}{\partial F(x_v)}.
\]
In the case of uniform margins $u_j=F_j(x_j)$, for a parametric copula cdf $C_{jv}(u_j,u_v)=
C_{jv}(u_j,u_v;\ftheta_{jv})$ this further simplifies to
\begin{equation}
h(u_j|u_v,\ftheta_{jv}):=\frac{\partial\,C_{j,v }(u_j,
u_v ;\ftheta_{jv})}{\partial u_v}.
\label{hfunc}
\end{equation}
With this notation we can express $F(x_j|\fx_{D})$ as
\begin{eqnarray*}
F(x_j|\fx_{D})
&=& h(F(x_j|\fx_{D_{-v}})|F(x_v|\fx_{D_{-v}}), \ftheta_{jv|D_{-v}}).
\end{eqnarray*}
This allows the recursive determination of the likelihood corresponding to
(\ref{dvine}). Furthermore, the inverse of the $h$-functions is used to facilitate
sampling from D- and C-vines (see for example \citealp{ACFB09} and
\citealp{kurowicka-csda-2007}). They are also used for sampling from the more general R-vine model
(see \citealp{stoeberczado2011}).

\subsection{D-vine based multivariate SCAR models}

We now combine bivariate SCAR models and D-vines to formulate a multivariate
D-vine SCAR model. For this we use a bivariate SCAR copula model as the pair copula
model in a D-vine copula. This gives rise to the following definition of a
D-vine SCAR copula density
\begin{equation}
\label{mscar-den}
c(u_1,\cdots,u_d; \bm \theta_t):=
\prod_{j=1}^{d-1} \prod_{i=1}^{d-j}
c(F(u_i|\bm u_{i+1:i+j-1};\bm \theta^{l(i,j)}_t),
F(u_{i+j}|\bm u_{i+1:i+j-1});\bm \theta^{l(i,j)}_t),
\end{equation}
where $l(i,j):=i,i+j|i+1:i+j-1$ and $\bm \theta_t:=
\{\bm \theta^{l(i,j)}_t ; j=1,\cdots, d-1,
i=1,\cdots,d-j\}$ is the time-varying copula parameter vector. Here
$c(\cdot,\cdot;\bm \theta^{l(i,j)}_t)$ is the bivariate copula
density corresponding to the bivariate SCAR copula given in (\ref{biscarden}),
where $\bm \theta^{l(i,j)}_t$ satisfies
\begin{equation}
\label{mscartheta}
\bm \theta^{l(i,j)}_t= r(\tau^{l(i,j)}_t)
=r(\psi(\lambda^{l(i,j)}_t))
\end{equation}
for the latent Gaussian AR(1) process $\lambda^{l(i,j)}_t$
with
\begin{equation}
\lambda^{l(i,j)}_t=\mu_{l(i,j)}+
\phi_{l(i,j)} (\lambda^{l(i,j)}_{t-1}
-\mu_{l(i,j)})+\sigma_{l(i,j)} z^{l(i,j)}_t.
\end{equation}
Here $z_{t}^{l(i,j)}$ are independent standard normal innovations for $j=1,\cdots, d;
i=1,\cdots,d-j$. As for the bivariate case, we assume
$|\phi_{l(i,j)}| <1$ and $\sigma_{l(i,j)} >0$ for stationarity and identification, respectively. The bivariate copula family
corresponding to $l(i,j)$ can be chosen arbitrarily and independent of any
other index $l(r,s)$.

The copula in (\ref{mscar-den}) can be used in (\ref{full-model})
to specify the joint distribution of the innovations in (\ref{ARMA-GARCH}).

\section{Parameter estimation in D-vine SCAR models}\label{Sec:Estimation}

We are interested in
estimating the parameters of both the marginal models and the stochastic copula
models. The joint density of our model is given by the product of the marginal
and the copula densities
\[
g(\varepsilon_{1,t},...,\varepsilon_{d,t})=c(F_i(\varepsilon_{1,t}),...,F_d(\varepsilon_{d,t}))\cdot
f_i(\varepsilon_{1,t})\cdot ... \cdot f_d(\varepsilon_{d,t}),
\]
where $g$, $c$ and $f$ denote the densities of the joint distribution, the
copula and the marginal distributions, respectively. Taking logarithms, we can
see that the joint log-likelihood is the sum the marginal and the copula
log-likelihood function. For estimation we ultilize a two-step approach common
in copula based models. In this approach first the marginal parameters are
estimated separately and standardized residuals are formed. These are
transformed using either a parametric (see \citealp{Joe2005}) or
nonparametric probability
integral transformation (see \citealp{Genest1995}) to get an independent sample from a
multivariate copula. These transformations do not change the dependence
structure among the standardized residuals. This approach allows us to
perform the estimation of the marginal and copula parameters separately. If the
marginal models are chosen carefully, as we will do, then a parametric probability
transformation is a good approximation to the true copula data
$u_{i,t}=F_i(\varepsilon_{i,t})$. Problems only occur if the marginal models are
grossly misspecified (see \citealp{Kim2007}).

Furthermore, we saw above that the density of a D-vine copula is the product of
bivariate (conditional) copulas. Therefore, instead of estimating all copula
parameters of our model in one step, which is computationally infeasible due to
the large number of parameters, we are able to estimate the copula parameters
sequentially.

In Section \ref{bi_SCAR_est} the estimation of bivariate SCAR copula models  by
simulated maximum likelihood (SML) using efficient importance sampling (EIS) is
reviewed, Section \ref{vine_est} presents the sequential estimation of vine
copula models and in Section \ref{MV_SCAR_est} we discuss how the sequential
estimation of D-vine SCAR copula models can be achieved.

\subsection{Estimation of bivariate SCAR copula models}\label{bi_SCAR_est}

For the moment, we are interested in estimating the copula parameter vector $\bm \omega:=(\mu, \phi, \sigma)$. For notational convenience we decided to
drop the indices $i$ and $j$ whenever no ambiguity arises. Denote  $\bm u_{i}=\{u_{i,t}\}_{t=1}^T$, $\bm u_{j}=\{u_{j,t}\}_{t=1}^T$ and $\bm
\Lambda=\{\lambda_{t}\}_{t=1}^T$ and let $f(\bm u_{i},\bm u_{j},\bm \Lambda;\bm \omega)$ be the joint density of the observable variables $(\bm u_i,\bm
u_j)$ and the latent process $\bm\Lambda$. Then the likelihood function of the parameter vector $\bm \omega$ can be obtained by integrating the latent
process $\bm \Lambda$ out of the joint likelihood,

\begin{align}
L(\bm \omega;\bm u_{i},\bm u_{j})=\int f(\bm u_{i},\bm u_{j},\bm \Lambda;\bm \omega)d \bm\Lambda.
\end{align}
We can alternatively write this as a product of conditional densities
\begin{equation}
L(\bm \omega;\bm u_{i},\bm u_{j})=\int \prod_{t=1}^T f(u_{i,t}, u_{j,t}, \lambda_t|\lambda_{t-1},\bm \omega) d \bm \Lambda.
\end{equation}

This is a T-dimensional integral that cannot be solved by analytical or numerical means. It can, however, be solved efficiently by Monte Carlo
integration using a technique called efficient importance sampling introduced by \citet{RZ07}. The idea is to make use of an auxiliary sampler
$m(\lambda_t;\lambda_{t-1},\bma_t)$ that utilizes the information on the latent process contained in the observable data. Note that it depends on the
auxiliary parameter vector $\bma_t=(a_{1,t},a_{2,t})$. Multiplying and dividing by $m(\cdot)$, the likelihood can then be rewritten as
\begin{equation}
\label{eq:EIS_LL_0}
L(\bm \omega;\bm u_{i},\bm u_{j})=\int \prod_{t=1}^T \left[\frac{f(u_{i,t}, u_{j,t}, \lambda_t|\lambda_{t-1},\bm \omega)}
{m(\lambda_t;\lambda_{t-1},\bma_t)}\right] \prod_{t=1}^T m(\lambda_t;\lambda_{t-1},\bma_t) d\bm \Lambda.
\end{equation}
Drawing $N$ trajectories $\tilde{\bm \Lambda}^{(i)}$ from the importance sampler\footnote{A good choice for $N$ is about 100.} the likelihood can be estimated by
\begin{equation}\label{eq:EIS_LL}
\tilde{L}(\bm \omega;\bm u_{i},\bm u_{j})=\frac{1}{N}\sum_{i=1}^N \left(\prod_{t=1}^T \left[\frac{f(u_{i,t}, u_{j,t},
 \tilde{\lambda}_t^{(i)}|\tilde{\lambda}_{t-1}^{(i)},\bm \omega)}{m(\tilde{\lambda}_t^{(i)};\tilde{\lambda}_{t-1}^{(i)},\bma_t)}\right]\right).
\end{equation}
This leaves the exact choice of the importance sampler $m(\cdot)$ to be determined, which ideally should provide a good
match between the numerator and the denominator of (\ref{eq:EIS_LL}) in order to minimize the variance of the likelihood function.
It is chosen to be
\begin{equation}
m(\lambda_t;\lambda_{t-1}, \bma_t) = \frac{k(\lambda_t, \lambda_{t-1};\bma_t)}{\chi(\lambda_{t-1};\bma_t)},
\end{equation}
where
\[
\chi(\lambda_{t-1};\bma_t) = \int k(\lambda_t, \lambda_{t-1};\bma_t) d \lambda_t
\]
is the normalizing constant of the auxiliary density kernel $k(\cdot)$. Furthermore, the choice
\[
k(\lambda_t, \lambda_{t-1};\bma_t) = p(\lambda_t| \lambda_{t-1},\bm \omega) \zeta(\lambda_t,\bma_t),
\]
with $p(\lambda_t|\lambda_{t-1},\bm \omega)$ the conditional density of $\lambda_t$ given $\lambda_{t-1}$ and
$\zeta(\lambda_t,\bma_t)= \exp(a_{1,t}\lambda_t + a_{2,t}\lambda_{t}^2)$ turns out to simplify the problem considerably. Noting that $f(u_{i,t}, u_{j,t}, \lambda_t|\lambda_{t-1},\bm \omega)=
c(u_{i,t}, u_{j,t}; \lambda_t) p(\lambda_t| \lambda_{t-1}, \bm \omega)$, the likelihood expression (\ref{eq:EIS_LL_0}) can be rewritten as
\begin{equation}\label{EIS_LL2}
L(\bm \omega;\bm u_{i},\bm u_{j})=\int \prod_{t=1}^T \left[\frac{c(u_{i,t},
u_{j,t}; \lambda_t)\chi(\lambda_{t};\bma_{t+1})}{\exp(a_{1,t}\lambda_t +
a_{2,t}\lambda_{t}^2)}\right] \prod_{t=1}^T m(\lambda_t;\lambda_{t-1},\bma_t) d\bm\Lambda,
\end{equation}
where we have used the fact that $\chi(\cdot)$ can be transferred back one period, because it does not depend on $\lambda_t$.
Defining $\chi(\lambda_T;\bma_{T+1}) \equiv 1$ and given a set of trajectories $\tilde{\Lambda}^{(i)}$ for $i=1,\ldots,N$, minimizing the sampling variance of the quotient in the likelihood function is equivalent to solving the following linear least squares problem for each period $t=T,\ldots,1$,
\begin{equation}\label{eq:LS_problem}
\log c(u_{i,t}, u_{j,t}; \tilde{\lambda}_t^{(i)}) + \log \chi(\lambda_{t};\bma_{t+1}) = c_t + a_{1,t} \tilde{\lambda}_t^{(i)} + a_{1,t} [\tilde{\lambda}_t^{(i)}]^2 + \eta_t^{(i)}.
\end{equation}
This problem can be solved by OLS with $c_t$ the regression intercept and $\eta_t^{(i)}$ the error term. Then the procedure works as follows. First, draw $N$ trajectories $\tilde{\Lambda}^{(i)}$ from $p(\lambda_t| \lambda_{t-1}, \omega)$ and estimate the auxiliary parameters $\hat{\bma}_t$ for $t=T,\ldots,1$ by solving (\ref{eq:LS_problem}). Next, draw $N$ trajectories $\tilde{\Lambda}^{(i)}$ from the importance sampler $m(\lambda_t;\lambda_{t-1}, \hat{\bma}_t)$ and re-estimate the auxiliary parameters $\{\hat{\bma}_t\}_{t=1}^T$. Iterate this procedure until convergence of $\{\hat{\bma}_t\}_{t=1}^T$ and use $N$ draws from the importance sampler to
 estimate the likelihood function (\ref{eq:EIS_LL}). This likelihood function
 can then be maximized to obtain parameter estimates $\hat{\bm \omega}$. Note that throughout the same random numbers have to be used in order to ensure convergence of $\{\hat{\bma}_t\}_{t=1}^T$ and smoothness of the likelihood function.

Although the parameter vector $\bm \omega$ driving the latent process is of some interest,
ultimately one wishes to get estimates of the latent process $\bm \Lambda$ and transformations thereof.
In particular, we are interested in estimating $\tau_t=\psi(\lambda_t)$ for $t=1,\ldots,T$, where $\psi(\cdot)$ denotes the inverse Fisher transform given in (\ref{eq:Inv_Fisher}). Smoothed estimates of $\psi(\lambda_t)$ given the entire history of the
observable information $\bm u_i$ and $\bm u_j$ can be computed as
\begin{equation}
\label{eq:EIS_psi}
E[\psi(\lambda_t)|\bm u_i,\bm u_j]=\frac{\int \psi(\lambda_t) f(\bm u_{i},\bm u_{j},\bm \Lambda;\bm \omega)d\bm \Lambda}{\int f(\bm u_{i},\bm u_{j},
\bm \Lambda;\omega)d\bm\Lambda}.
\end{equation}
Note that the denominator in (\ref{eq:EIS_psi}) corresponds to the likelihood function and
both integrals can be estimated using draws from the importance sampler $m(\lambda_t;\lambda_{t-1}, \hat{\bma}_t)$. Filtered estimates of $\psi(\lambda_t)$ given information
until time $t-1$ can be computed in a similar way and details are given in \citet{LR03}.

\subsection{Sequential estimation of D-vine copula parameters}\label{vine_est}
The form of the D-vine density given in (\ref{dvine}) allows for a
sequential parameter estimation approach starting from the first tree until
the last tree. This was first proposed by \citet{ACFB09} for D-vines and shown in detail for C-vines in
\citet{schepsmeier}. First estimate the parameters corresponding to the pair-copulas
in the first tree using any method you prefer. For the copula parameters
identified in the second tree, one first has to transform the data with the
$h$ function in (\ref{hfunc}) required for the appropriate conditional cdf using estimated
parameters to determine pseudo realizations needed in the second tree. Using these pseudo observations the
parameters in the second tree are estimated, the pseudo data is again transformed using the $h$ function and so on.

For example we want to estimate the parameters of copula $c_{{13|2}}$. First
transform the observations $\{u_{1,t},u_{2,t},u_{3,t}, t=1,\cdots,n\}$ to
$u_{1|2,t}:=h(u_{1,t}|u_{2,t},\hat{\theta}_{12})$ and
$u_{3|2,t}:=h(u_{3,t}|u_{2,t},\hat{\theta}_{23})$, where $\hat{\theta}_{12}$
and $\hat{\theta}_{23}$ are the estimated parameters in the first tree.   Now
estimate $\theta_{13|2}$ based on $\{u_{1|2,t},u_{3|2,t}; t=1,\cdots,n\}$.
Continue sequentially until all copula parameters of all trees are estimated.
For trees $T_i$ with $i \geq 2 $ recursive applications of the $h$ functions
is needed. Asymptotic normality of the SE has been established by
\citet{haff}. However, the asymptotic covariance of the parameter estimates is very complex and one has to
resort to bootstrapping to estimate the standard errors. SE is often used
in large dimensional problems, e.g. \citet{mendes}, \citet{heinen:2008},
\citet{brechmann-etal} and \citet{BrechmannCzado2011}. A joint MLE of all
parameters in (\ref{dvine}) requires high dimensional optimization. Therefore
SE's are often used as starting values as, e.g., in \citet{ACFB09} and
\citet{schepsmeier}.\footnote{From a practical perspective, the recent R package {\rm  CDVine} of \citet{CDVine}
provides easy to use random number generation, and both SE and MLE fitting algorithms for C- and
D-vines.}

A final issue to be discussed is model selection for D-vine copula models.
For non
Gaussian pair copulas, permutations of the ordering of the variables give
different D-vine copulas. In fact, there are $d!/2$ different D-vine copulas
when a common bivariate copula is used as pair copula type. Often the bivariate
Clayton, Gumbel, Gauss, t, Joe and Frank copula families are utilized as
choices for pair copula terms. However, in this study we restrict the attention
to the Gauss, Gumbel, Clayton and rotated versions thereof.

\subsection{Estimation of D-vine SCAR models}\label{MV_SCAR_est}
In principle, estimation of the D-vine SCAR model given in (\ref{mscar-den}) works the same way as for static D-vine
copulas. There are, however, two important differences. First of all, given that the
bivariate SCAR models in the first tree have been estimated, it is not possible to
apply the $h$ function given in (\ref{hfunc}) directly to obtain the pseudo
observations that are needed to obtain the parameters on the second tree. The
reason is that one only obtains parameter estimates of the hyper-parameters $(\mu, \phi, \sigma)$,
but not of the latent (time-varying) copula parameters $\theta_t$. We do, however, have $N$
simulated trajectories $\tilde{\theta}^{(i)}_t$ from the importance sampler. With these we can calculate the pseudo observations by
\begin{equation}
u_{j|v,t}=\frac{1}{N}\sum_{i=1}^N h(u_{j,t}|u_{v,t}, \tilde{\theta}_t^{(i)}),
\end{equation}
where we suppress the dependence of $\tilde{\theta}$ on the variable indices $j$ and $v$ for notational reasons.\footnote{Alternatively, we
 could calculate the
pseudo realizations using the smoothed estimates of the latent dependence parameter using (\ref{eq:EIS_psi}). However, averaging over the nonlinear  transformation $h$ seems more
reasonable than applying the transformation to the (weighted) average.} The second difference to the static D-vine copula model is that one-step estimation
by MLE is computationally not feasible for the time-varying model, because each bivariate likelihood function needs to be computed by simulation.

For a $d$ dimensional dataset the D-vine SCAR copula one has to estimate $3d(d+1)/2$ parameters. Fortunately, we can reduce the number
of parameters to be estimated by placing a number of restrictions. Similar to tail properties of D-vines studied in \citet{joe2010} the choice of
time-varying pair copulas in the first tree propagates to the whole
distribution, in particular all pairs of variables have an induced time-varying
Kendall's tau. We expect estimation errors to increase for parameters
as the corresponding tree increases because of the sequential nature of the
estimation procedure. Therefore we allow for D-vine SCAR copula
models where the pair copulas are time-varying only in lower trees, while the
pair copulas are time-constant for higher trees. Such models will also be
investigated in our simulation study.

A second useful restriction is to allow for the possibility of truncating the D-vine copula, which means that we set all pair copulas beyond
a certain tree equal to the independence copula. This is empirically justified, since the dependence in the lower trees seems to capture most of
the overall dependence in the data and the conditional dependence in higher trees is hardly visible. Note that this also allows the estimation of our model in arbitrarily large dimensions,
as we will only need to estimate (bivariate) models up to a certain dimension and can truncate the model thereafter. For static models this has been followed by \citet{brechmann-etal} and includes tests at which level to truncate.

In order to decide which copula family to use and whether to use time-varying, time-constant or independence copulas at certain levels we compare the Bayesian Information Criterion (BIC)
for all competing models. We decided for this information criterion, because it favors parsimonious models. Given the high flexibility of the D-vine SCAR model and the difficulty to
estimate the parameters at higher level, we believe that parsimony is crucial.

\subsection{Computational issues}
Estimation of a bivariate SCAR model by simulated maximum likelihood programmed in MATLAB can take up to several minutes on a normal computer. As we consider a Monte Carlo simulation
and an application with 29 variables this is much too slow. However, the problem at hand offers itself to parallel computing. On each tree one has to estimate a large
number of bivariate models independently of each other. Therefore, given a sufficient number of processing cores we can estimate all models on one tree at the same time and proceed to the next
tree once all models are estimated. Depending on the dimension of the problem, this can lead to immense increases in computing speed. The computations
for the Monte Carlo study in the next section and our empirical application were performed on a Linux cluster computer computer with 32 processing cores (Quad Core AMD  Opteron, 2.6Mhz),
The most demanding computational task, the estimation of the log-likelihood function by EIS, is implemented in C++, which resulted in our code being about 20-30 times faster compared to MATLAB code. The maximization of the likelihood and the parallel computation within levels is implemented in R (version 2.12.1) by using the \texttt{optim} function and the multicore library.

\section{Simulation study}\label{Sec:Simulations}

To investigate the performance of the sequential estimation utilizing EIS for
D-vine SCAR copula models we conducted an extensive simulation study.\footnote{For brevity we only present a part of the results and the outcomes for further
parameter and model constellations are available from the authors upon request.} For this we chose a
four dimensional setup. We simulated 1000 four dimensional SCAR D-vine copula
data sets of length 1000 under several scenarios. Then we estimated the relative bias
and relative MSE for the parameters of the latent AR(1) processes corresponding
to each of the six bivariate copula terms of the D-vine copula. Standard errors are
also estimated and given in parentheses in the tables.

We expect that the stationary latent AR(1) signal-to-noise ratio given by $sn:=\frac{\mu}{\sigma (1-\phi^2)^{-1/2}}$ and
 the stationary variance
$avar:= \frac{\sigma^2}{1-\phi^2}$ will influence the performance. Therefore we include these quantities as well.
The copula families we consider are the Gaussian, Clayton and Gumbel copulas.

In the following we present the small sample performance results for scenarios involving
a single bivariate copula family for all pair copula terms and a common
time-varying parameter structure for all pair copula terms (Section \ref{Sec:sim1}),
a single bivariate copula family for all pair copula terms and a common
time-varying parameter structure for only  pair copula terms in the first tree (Section \ref{Sec:sim2}), and
different time-varying structures for all or only the first tree pair copulas and/or different copula families (Section \ref{Sec:sim3}).
In Section \ref{Sec:sim4} we draw conclusions from our simulation study relevant for the subsequent application of our model.

\subsection{Common copula families and common time-varying structure for all pair copulas}\label{Sec:sim1}

The results in Table \ref{sim-common} show satisfactory results for all latent
AR(1) parameters of unconditional pair copula terms, i.e. the terms with
indices $12, 23, 34$. The mean parameters $\mu$ and the standard error
parameters $\sigma$ of the pair copulas in trees 2 to 3 are also well estimated in
scenarios where the asymptotic AR(1) signal-to-noise ratio $sn$ is large.  The
estimation of the persistence parameter $\phi$ is less effected by the value of
the signal-to-noise ratio. By the sequential nature of the estimation procedure
we expect the performance
to deteriorate for parameters with increasing conditioning set. This behavior is
also visible. In particular, the estimation on the second tree is still reasonably precise for most cases, but on the third level the results become significantly worse. Comparing
the results across different copula families one can generally state that for the Gaussian copula the results are best, closely
followed by the Gumbel copula. For the Clayton copula the estimation is most imprecise both in terms of relative bias and MSE.

\begin{table}[t!]
{\scriptsize
  \begin{center}
  \begin{tabular}{cccc|r@{(}l@{)\ \ } r@{(}l@{)\ \  } r@{(}l@{)\ \  } |  r@{(}r@{)\ \  } r@{(}r@{)\ \  } r@{(}r@{)\ } | c c}
\hline
    Index & \multicolumn{3}{c|}{True values} & \multicolumn{6}{c|}{Average relative Bias} & \multicolumn{6}{c|}{Average relative MSE}
& sn & avar\\
    & $\mu$  & $\phi$ & $\sigma$ &\multicolumn{2}{c}{$\mu$}  & \multicolumn{2}{c}{$\phi$} & \multicolumn{2}{c|}{$\sigma$}  & \multicolumn{2}{c}{$\mu$ } &     \multicolumn{2}{c}{$\phi$}     & \multicolumn{2}{c|}{$\sigma$ }  \\


   \MR{ Copula families:  Gaussian Gaussian Gaussian Gaussian Gaussian Gaussian  }12 & .50 & .95 & .15 & .0298 & .0071 & -.0116 & .0008 & -.0761 & .0033 & .0503 & .0041 & .0008 & .0001 & .0166 & .0009 & 1.04 & 0.23 \\
  23 & .50 & .95 & .15 & .0361 & .0075 & -.0123 & .0009 & -.0743 & .0035 & .0565 & .0048 & .0009 & .0001 & .0179 & .0011 & 1.04 & 0.23 \\
  34 & .50 & .95 & .15 & .0335 & .0077 & -.0108 & .0008 & -.0800 & .0034 & .0603 & .0059 & .0008 & .0001 & .0176 & .0008 & 1.04 & 0.23 \\
  13$|$2 & .50 & .95 & .15 & -.1036 & .0051 & -.0061 & .0006 & -.2112 & .0036 & .0362 & .0016 & .0004 & .0000 & .0573 & .0016 & 1.04 & 0.23 \\
  24$|$3 & .50 & .95 & .15 & -.0997 & .0054 & -.0064 & .0006 & -.2154 & .0034 & .0385 & .0016 & .0003 & .0000 & .0581 & .0016 & 1.04 & 0.23 \\
  14$|$23 & .50 & .95 & .15 & -.4090 & .0042 & -.0215 & .0011 & -.3230 & .0050 & .1846 & .0035 & .0017 & .0001 & .1286 & .0031 & 1.04 & 0.23 \\
   \MR{ Copula families :  Clayton Clayton Clayton Clayton Clayton Clayton  }12 & .50 & .95 & .15 & .0580 & .0084 & -.0092 & .0011 & -.0372 & .0040 & .0682 & .0103 & .0012 & .0001 & .0162 & .0011 & 1.04 & 0.23 \\
  23 & .50 & .95 & .15 & .0743 & .0085 & -.0096 & .0011 & -.0384 & .0041 & .0723 & .0095 & .0013 & .0002 & .0172 & .0014 & 1.04 & 0.23 \\
  34 & .50 & .95 & .15 & .0592 & .0084 & -.0086 & .0010 & -.0394 & .0040 & .0684 & .0094 & .0011 & .0001 & .0165 & .0012 & 1.04 & 0.23 \\
  13$|$2 & .50 & .95 & .15 & -.1915 & .0047 & -.0110 & .0008 & -.2585 & .0040 & .0572 & .0022 & .0007 & .0001 & .0818 & .0022 & 1.04 & 0.23 \\
  24$|$3 & .50 & .95 & .15 & -.1956 & .0048 & -.0115 & .0007 & -.2615 & .0038 & .0596 & .0022 & .0006 & .0000 & .0816 & .0021 & 1.04 & 0.23 \\
  14$|$23 & .50 & .95 & .15 & -.5676 & .0036 & -.0379 & .0019 & -.4283 & .0059 & .3338 & .0039 & .0048 & .0007 & .2148 & .0046 & 1.04 & 0.23 \\
   \MR{Copula families :  Gumbel Gumbel Gumbel Gumbel Gumbel Gumbel  }12 & .50 & .95 & .15 & .0210 & .0063 & -.0036 & .0006 & -.0220 & .0036 & .0393 & .0024 & .0004 & .0001 & .0133 & .0007 & 1.04 & 0.23 \\
  23 & .50 & .95 & .15 & .0258 & .0064 & -.0045 & .0006 & -.0238 & .0039 & .0417 & .0025 & .0004 & .0001 & .0154 & .0015 & 1.04 & 0.23 \\
  34 & .50 & .95 & .15 & .0142 & .0064 & -.0038 & .0006 & -.0267 & .0037 & .0409 & .0032 & .0004 & .0000 & .0143 & .0006 & 1.04 & 0.23 \\
  13$|$2 & .50 & .95 & .15 & -.1196 & .0053 & -.0045 & .0006 & -.2187 & .0040 & .0424 & .0024 & .0004 & .0000 & .0635 & .0018 & 1.04 & 0.23 \\
  24$|$3 & .50 & .95 & .15 & -.1168 & .0052 & -.0050 & .0006 & -.2217 & .0038 & .0405 & .0018 & .0004 & .0000 & .0633 & .0018 & 1.04 & 0.23 \\
  14$|$23 & .50 & .95 & .15 & -.4553 & .0040 & -.0220 & .0012 & -.3703 & .0052 & .2234 & .0036 & .0020 & .0002 & .1643 & .0037 & 1.04 & 0.23 \\

   \MR{ Copula families :  Gaussian Gaussian Gaussian Gaussian Gaussian Gaussian  }12 & .50 & .95 & .05 & -.0023 & .0022 & -.0177 & .0017 & .0410 & .0112 & .0050 & .0003 & .0033 & .0005 & .1325 & .0097 & 3.12 & 0.03 \\
  23 & .50 & .95 & .05 & .0014 & .0023 & -.0169 & .0020 & .0459 & .0107 & .0055 & .0003 & .0043 & .0021 & .1227 & .0081 & 3.12 & 0.03 \\
  34 & .50 & .95 & .05 & .0029 & .0023 & -.0164 & .0016 & .0506 & .0109 & .0057 & .0003 & .0030 & .0005 & .1274 & .0105 & 3.12 & 0.03 \\
  13$|$2 & .50 & .95 & .05 & .0057 & .0023 & -.0192 & .0022 & .0704 & .0111 & .0058 & .0003 & .0053 & .0031 & .1354 & .0081 & 3.12 & 0.03 \\
  24$|$3 & .50 & .95 & .05 & .0013 & .0022 & -.0214 & .0020 & .0692 & .0123 & .0053 & .0003 & .0047 & .0010 & .1636 & .0144 & 3.12 & 0.03 \\
  14$|$23 & .50 & .95 & .05 & -.0740 & .0023 & -.0310 & .0029 & .0417 & .0138 & .0109 & .0006 & .0100 & .0036 & .2024 & .0154 & 3.12 & 0.03 \\
   \MR{ Copula families :  Clayton Clayton Clayton Clayton Clayton Clayton  }12 & .50 & .95 & .05 & -.0028 & .0022 & -.0163 & .0016 & .0465 & .0105 & .0049 & .0003 & .0028 & .0005 & .1161 & .0086 & 3.12 & 0.03 \\
  23 & .50 & .95 & .05 & .0011 & .0022 & -.0155 & .0014 & .0586 & .0104 & .0052 & .0002 & .0021 & .0003 & .1170 & .0091 & 3.12 & 0.03 \\
  34 & .50 & .95 & .05 & .0030 & .0023 & -.0141 & .0012 & .0531 & .0098 & .0054 & .0003 & .0017 & .0002 & .1029 & .0067 & 3.12 & 0.03 \\
  13$|$2 & .50 & .95 & .05 & -.0365 & .0021 & -.0162 & .0013 & -.0321 & .0099 & .0060 & .0003 & .0021 & .0002 & .1037 & .0061 & 3.12 & 0.03 \\
  24$|$3 & .50 & .95 & .05 & -.0415 & .0021 & -.0157 & .0014 & -.0528 & .0103 & .0061 & .0003 & .0024 & .0004 & .1134 & .0069 & 3.12 & 0.03 \\
  14$|$23 & .50 & .95 & .05 & -.2252 & .0019 & -.0727 & .0043 & .1145 & .0175 & .0544 & .0009 & .0249 & .0043 & .3317 & .0199 & 3.12 & 0.03 \\
   \MR{ Copula families :  Gumbel Gumbel Gumbel Gumbel Gumbel Gumbel  }12 & .50 & .95 & .05 & -.0021 & .0023 & -.0227 & .0021 & .0929 & .0139 & .0053 & .0003 & .0050 & .0009 & .2073 & .0182 & 3.12 & 0.03 \\
  23 & .50 & .95 & .05 & .0015 & .0024 & -.0237 & .0026 & .1005 & .0136 & .0060 & .0004 & .0075 & .0024 & .2022 & .0198 & 3.12 & 0.03 \\
  34 & .50 & .95 & .05 & .0020 & .0025 & -.0222 & .0022 & .1045 & .0137 & .0063 & .0006 & .0054 & .0013 & .2043 & .0170 & 3.12 & 0.03 \\
  13$|$2 & .50 & .95 & .05 & -.0084 & .0024 & -.0190 & .0017 & .0459 & .0127 & .0062 & .0004 & .0032 & .0005 & .1695 & .0131 & 3.12 & 0.03 \\
  24$|$3 & .50 & .95 & .05 & -.0131 & .0023 & -.0205 & .0021 & .0332 & .0137 & .0056 & .0003 & .0048 & .0011 & .1940 & .0177 & 3.12 & 0.03 \\
  14$|$23 & .50 & .95 & .05 & -.1324 & .0022 & -.0245 & .0021 & -.0670 & .0137 & .0227 & .0008 & .0053 & .0013 & .1980 & .0145 & 3.12 & 0.03 \\ \hline
\end{tabular}
\caption{Estimated relative bias and relative MSE with estimated standard errors for
four dimensional D-vine SCAR copulas with common bivariate copula
family and common time-varying structure for all pair copulas.}
\label{sim-common}
\end{center}
}
\end{table}

\subsection{Common copula families and common time-varying structure for pair copula terms in the first tree only}\label{Sec:sim2}

The small sample performance for scenarios where only the copula parameter for pair copulas in the
first tree is time-varying, reported in Table \ref{sim-common-first}, is quite satisfactory for all models except for the mean parameters $\mu$ in the third
tree. This was to be expected by the results from the previous section. Note, however, that mean dependence, measured by the parameter $\mu$, in the second and third
tree is lower than in the above scenario. We can therefore conclude that the precision in estimating $\mu$ decreases when the degree of dependence decreases. When the overall dependence is
comparable, as is the case for the pair 13$|$2 in the second tree, we see that the parameter $\mu$ is estimated much more precisely than for the
time-varying case. Furthermore, note that the average relative bias on the higher trees is positive for the copulas with lower $\mu$ parameter.

\begin{table}[t!]
{\scriptsize
  \begin{center}
  \begin{tabular}{cccc|r@{(}l@{)\ \ } r@{(}l@{)\ \  } r@{(}l@{)\ \  } |  r@{(}r@{)\ \  } r@{(}r@{)\ \  } r@{(}r@{)\ } | c c}
\hline
    Ind & \multicolumn{3}{c|}{True values} & \multicolumn{6}{c|}{Average relative Bias} & \multicolumn{6}{c|}{Average relative MSE}  &
    sn & avar\\
    & $\mu$  & $\phi$ & $\sigma$ &\multicolumn{2}{c}{$\mu$}  & \multicolumn{2}{c}{$\phi$} & \multicolumn{2}{c|}{$\sigma$}  & \multicolumn{2}{c}{$\mu$ } &     \multicolumn{2}{c}{$\phi$}     & \multicolumn{2}{c|}{$\sigma$ }  \\

%
   \MR{Copula families:  Gaussian Gaussian Gaussian Gaussian Gaussian Gaussian  }12 & .50 & .95 & .15 & .0395 & .0083 & -.0109 & .0009 & -.0827 & .0035 & .0689 & .0075 & .0009 & .0001 & .0192 & .0011 & 1.04 & 0.23 \\
  23 & .50 & .95 & .15 & .0291 & .0070 & -.0127 & .0009 & -.0721 & .0033 & .0493 & .0034 & .0010 & .0001 & .0158 & .0007 & 1.04 & 0.23 \\
  34 & .50 & .95 & .15 & .0263 & .0078 & -.0113 & .0008 & -.0763 & .0034 & .0598 & .0055 & .0008 & .0001 & .0171 & .0009 & 1.04 & 0.23 \\
  13$|$2 & .50 & .00 & .00 & -.0156 & .0012 & - & - & - & - & .0016 & .0001 & - & - & - & - & Inf & 0 \\
  24$|$3 & .30 & .00 & .00 & .0416 & .0022 & - & - & - & - & .0067 & .0003 & - & - & - & - & Inf & 0 \\
  14$|$23 & .20 & .00 & .00 & .2185 & .0039 & - & - & - & - & .0624 & .0017 & - & - & - & - & Inf & 0 \\
   \MR{Copula families :  Clayton Clayton Clayton Gaussian Gaussian Gaussian  }12 & .50 & .95 & .15 & .0781 & .0088 & -.0077 & .0011 & -.0403 & .0041 & .0760 & .0095 & .0011 & .0002 & .0166 & .0011 & 1.04 & 0.23 \\
  23 & .50 & .95 & .15 & .0712 & .0088 & -.0110 & .0012 & -.0326 & .0041 & .0751 & .0115 & .0014 & .0002 & .0162 & .0012 & 1.04 & 0.23 \\
  34 & .50 & .95 & .15 & .0544 & .0091 & -.0091 & .0011 & -.0347 & .0045 & .0777 & .0103 & .0012 & .0002 & .0192 & .0014 & 1.04 & 0.23 \\
  13$|$2 & .50 & .00 & .00 & -.0152 & .0012 & - & - & - & - & .0015 & .0001 & - & - & - & - & Inf & 0 \\
  24$|$3 & .30 & .00 & .00 & .0424 & .0023 & - & - & - & - & .0067 & .0003 & - & - & - & - & Inf & 0 \\
  14$|$23 & .20 & .00 & .00 & .2237 & .0043 & - & - & - & - & .0671 & .0020 & - & - & - & - & Inf & 0 \\
   \MR{Copula families :  Gumbel Gumbel Gumbel Gaussian Gaussian Gaussian  }12 & .50 & .95 & .15 & .0282 & .0063 & -.0043 & .0007 & -.0259 & .0038 & .0404 & .0022 & .0005 & .0001 & .0153 & .0007 & 1.04 & 0.23 \\
  23 & .50 & .95 & .15 & .0243 & .0065 & -.0055 & .0007 & -.0193 & .0040 & .0421 & .0023 & .0005 & .0001 & .0161 & .0010 & 1.04 & 0.23 \\
  34 & .50 & .95 & .15 & .0114 & .0069 & -.0039 & .0006 & -.0251 & .0039 & .0473 & .0055 & .0004 & .0000 & .0157 & .0013 & 1.04 & 0.23 \\
  13$|$2 & .50 & .00 & .00 & -.0070 & .0011 & - & - & - & - & .0013 & .0001 & - & - & - & - & Inf & 0 \\
  24$|$3 & .30 & .00 & .00 & .0500 & .0022 & - & - & - & - & .0074 & .0003 & - & - & - & - & Inf & 0 \\
  14$|$23 & .20 & .00 & .00 & .2346 & .0038 & - & - & - & - & .0696 & .0018 & - & - & - & - & Inf & 0 \\

   \MR{Copula families:  Gaussian Gaussian Gaussian Gaussian Gaussian Gaussian  }12 & .50 & .95 & .05 & -.0012 & .0025 & -.0168 & .0020 & .0465 & .0112 & .0064 & .0007 & .0045 & .0024 & .1331 & .0105 & 3.12 & 0.03 \\
  23 & .50 & .95 & .05 & .0032 & .0023 & -.0179 & .0017 & .0575 & .0108 & .0054 & .0003 & .0034 & .0009 & .1266 & .0087 & 3.12 & 0.03 \\
  34 & .50 & .95 & .05 & .0037 & .0022 & -.0160 & .0018 & .0455 & .0106 & .0052 & .0002 & .0036 & .0014 & .1191 & .0079 & 3.12 & 0.03 \\
  13$|$2 & .50 & .00 & .00 & .0059 & .0012 & - & - & - & - & .0015 & .0001 & - & - & - & - & Inf & 0 \\
  24$|$3 & .30 & .00 & .00 & .0295 & .0021 & - & - & - & - & .0053 & .0002 & - & - & - & - & Inf & 0 \\
  14$|$23 & .20 & .00 & .00 & .2502 & .0033 & - & - & - & - & .0740 & .0017 & - & - & - & - & Inf & 0 \\
   \MR{Copula families:  Clayton Clayton Clayton Gaussian Gaussian Gaussian  }12 & .50 & .95 & .05 & -.0014 & .0024 & -.0191 & .0023 & .0631 & .0107 & .0060 & .0008 & .0060 & .0023 & .1233 & .0132 & 3.12 & 0.03 \\
  23 & .50 & .95 & .05 & .0020 & .0022 & -.0155 & .0013 & .0573 & .0099 & .0051 & .0002 & .0019 & .0002 & .1061 & .0068 & 3.12 & 0.03 \\
  34 & .50 & .95 & .05 & .0041 & .0022 & -.0135 & .0012 & .0461 & .0096 & .0052 & .0002 & .0016 & .0002 & .0994 & .0070 & 3.12 & 0.03 \\
  13$|$2 & .50 & .00 & .00 & .0061 & .0012 & - & - & - & - & .0015 & .0001 & - & - & - & - & Inf & 0 \\
  24$|$3 & .30 & .00 & .00 & .0284 & .0021 & - & - & - & - & .0052 & .0002 & - & - & - & - & Inf & 0 \\
  14$|$23 & .20 & .00 & .00 & .2482 & .0034 & - & - & - & - & .0740 & .0018 & - & - & - & - & Inf & 0 \\
   \MR{Copula families :  Gumbel Gumbel Gumbel Gaussian Gaussian Gaussian  }12 & .50 & .95 & .05 & -.0010 & .0026 & -.0211 & .0022 & .0902 & .0136 & .0069 & .0007 & .0056 & .0023 & .2006 & .0163 & 3.12 & 0.03 \\
  23 & .50 & .95 & .05 & .0036 & .0023 & -.0247 & .0026 & .1116 & .0135 & .0054 & .0002 & .0079 & .0031 & .2035 & .0144 & 3.12 & 0.03 \\
  34 & .50 & .95 & .05 & .0044 & .0024 & -.0223 & .0023 & .0964 & .0133 & .0058 & .0004 & .0061 & .0020 & .1935 & .0161 & 3.12 & 0.03 \\
  13$|$2 & .50 & .00 & .00 & .0042 & .0012 & - & - & - & - & .0014 & .0001 & - & - & - & - & Inf & 0 \\
  24$|$3 & .30 & .00 & .00 & .0262 & .0021 & - & - & - & - & .0051 & .0002 & - & - & - & - & Inf & 0 \\
  14$|$23 & .20 & .00 & .00 & .2472 & .0034 & - & - & - & - & .0730 & .0018 & - & - & - & - & Inf & 0 \\ \hline
\end{tabular}
\caption{Estimated relative bias and relative MSE with estimated standard errors for
 four dimensional D-vine SCAR copulas with common bivariate copula
family and common time-varying structure for all first tree pair copulas.}
\label{sim-common-first}

\end{center}
}
\end{table}

\subsection{Different time-varying structure and mixed or common pair copula families}\label{Sec:sim3}
Since many scenarios are possible in this setup, we restrict ourselves to
a three dimensional setup assuming all pair copulas time-varying with common pair copula family or different pair copula families.
In particular, we vary the persistence parameter, which is now lower for one copula on the first tree and the copula on the second tree.
Again we report estimated relative bias and relative MSE together with the signal-to-noise and stationary variance of the latent AR(1) process
in Table \ref{sim-mixed}.
As before, the results are based on 1000 simulated data sets of length 1000.

The results in Table \ref{sim-mixed} show only a moderate influence of different time-varying structures or different pair copula families. It is
notable that lower persistence leads to more precise estimation of $\mu$ and $\phi$ and less precision for $\sigma$, but that effect can mainly be
explained by the changed signal-to-noise ratio.
Again parameters of the higher tree are less well estimated compared to the ones in the first tree.

\begin{table}[t!]
{\scriptsize
  \begin{center}
  \begin{tabular}{cccc|r@{(}l@{)\ \ } r@{(}l@{)\ \  } r@{(}l@{)\ \  } |  r@{(}r@{)\ \  } r@{(}r@{)\ \  } r@{(}r@{)\ } | c c}
\hline
    index & \multicolumn{3}{c|}{True values} & \multicolumn{6}{c|}{relative bias} & \multicolumn{6}{c|}{relative MSE}  &
 sn & avar\\
    & $\mu$  & $\phi$ & $\sigma$ &\multicolumn{2}{c}{$\mu$}  & \multicolumn{2}{c}{$\phi$} & \multicolumn{2}{c|}{$\sigma$}  &
     \multicolumn{2}{c}{$\mu$ } &     \multicolumn{2}{c}{$\phi$}     & \multicolumn{2}{c|}{$\sigma$ }  \\

  \MR{ Copula families:  Gaussian Gaussian Gaussian  }12 & .50 & .85 & .15 & -.0064 & .0023 & .0035 & .0015 & -.0864 & .0045 & .0057 & .0003 & .0023 & .0001 & .0280 & .0013 & 1.76 & 0.08 \\
  23 & .50 & .95 & .15 & .0403 & .0073 & -.0117 & .0009 & -.0774 & .0035 & .0567 & .0050 & .0009 & .0001 & .0190 & .0010 & 1.04 & 0.23 \\
  13$|$2 & .50 & .85 & .15 & -.0379 & .0021 & -.0022 & .0020 & -.1801 & .0049 & .0059 & .0003 & .0041 & .0003 & .0577 & .0020 & 1.76 & 0.08 \\
   \MR{Copula families  :  Clayton Clayton Clayton  }12 & .50 & .85 & .15 & -.0026 & .0023 & -.0014 & .0014 & -.0541 & .0044 & .0053 & .0002 & .0021 & .0001 & .0221 & .0010 & 1.76 & 0.08 \\
  23 & .50 & .95 & .15 & .0696 & .0075 & -.0091 & .0010 & -.0335 & .0041 & .0608 & .0079 & .0011 & .0002 & .0179 & .0015 & 1.04 & 0.23 \\
  13$|$2 & .50 & .85 & .15 & -.1193 & .0018 & -.0067 & .0022 & -.2900 & .0049 & .0176 & .0005 & .0048 & .0003 & .1085 & .0029 & 1.76 & 0.08 \\
   \MR{Copula families:  Gumbel Gumbel Gumbel  }12 & .50 & .85 & .15 & -.0034 & .0024 & -.0053 & .0017 & -.0410 & .0054 & .0058 & .0003 & .0030 & .0002 & .0312 & .0014 & 1.76 & 0.08 \\
  23 & .50 & .95 & .15 & .0245 & .0061 & -.0038 & .0006 & -.0254 & .0037 & .0393 & .0021 & .0004 & .0001 & .0145 & .0008 & 1.04 & 0.23 \\
  13$|$2 & .50 & .85 & .15 & -.0588 & .0020 & .0036 & .0020 & -.2254 & .0054 & .0078 & .0003 & .0043 & .0003 & .0806 & .0026 & 1.76 & 0.08 \\
   \MR{Copula families  :  Clayton Gumbel Gaussian  }12 & .50 & .85 & .15 & -.0018 & .0023 & -.0013 & .0014 & -.0545 & .0043 & .0053 & .0002 & .0021 & .0001 & .0222 & .0010 & 1.76 & 0.08 \\
  23 & .50 & .95 & .15 & .0290 & .0064 & -.0031 & .0006 & -.0270 & .0037 & .0430 & .0025 & .0004 & .0001 & .0146 & .0007 & 1.04 & 0.23 \\
  13$|$2 & .50 & .85 & .15 & -.0372 & .0020 & -.0038 & .0022 & -.1854 & .0050 & .0057 & .0002 & .0049 & .0005 & .0606 & .0020 & 1.76 & 0.08 \\

\end{tabular}
\caption{Estimated relative bias and relative MSE together with estimated standard errors in parentheses for selected senarios of three dimensional
SCAR D-vine copulas assuming different time-varying structures and common or different
pair copula families.}
\label{sim-mixed}
\end{center}
}
\end{table}

\subsection{Conclusions from the simulation study}\label{Sec:sim4}
Overall we saw that the estimation becomes worse on higher trees and that this effect is stronger when we allow for time variation. The relative imprecision also
increases with lower overall dependence. For our application we can conclude that it may be sensible to restrict the time variation to the first few trees and
restrict the copula parameter to be constant beyond. Furthermore, placing the pairs with the largest dependence on the first tree as is commonly done
for D-vine copula models is expected to provide the most precise estimates for the time variation in dependence, which is what we are mainly interested
in here.

\section{Application}\label{Sec:Application}

In this section we provide an empirical illustration of the D-vine SCAR model. The dataset we consider are daily returns from 29 stocks listed in the
DAX30 index during the period from the 1$^{st}$ of January 2007 to the 14$^{th}$ of June 2011, giving a total of 1124 observations. A list of the
included companies is given in the Appendix. We decided for this dataset to find a balance between demonstrating the possibility of high dimensional
modeling and the ability to still present the main results. Nevertheless, one could in principle consider much larger dimensions, which we leave for
future research.

The conditional mean of the returns was modeled using an ARMA(1,1) model for all stock separately. The Ljung-Box statistics for the residuals revealed no
significant remaining autocorrelation. For the conditional variance we considered GARCH(1,1) models with Student t innovations. Although one may
argue that GARCH models that allow for the leverage effect such as the GJR-GARCH are appropriate for many individuals stocks, preliminary results
suggested that the results for the dependence modeling are not affected by this choice. Results for the marginal models are not reported here for
brevity but are available upon request.

Next, we estimated the D-vine SCAR model on the transformed standardized residuals. The ordering of the variables was done by maximizing the
overall pairwise dependence measured by Kendall's tau. In particular, first choose the pair of variables with the highest empirical Kendall's $\tau$. Second connect the
next
variable which has highest pairwise Kendall's $\tau$ with one of the previously chosen variables and proceed in a similar fashion until all variables are
connected. This is the common strategy for D-vine copulas and is also motivated by our simulation results.
In particular, we expect to capture the overall time variation of the dependence as good as possible with this choice, as it turns out that time variation
is most relevant on the first tree. For each bivariate (conditional) copula model we then face two important choices, namely whether dependence is time-varying
or static, and which copula family to use. We automatically select the model by first estimating time-varying and constant copulas from the following families: Gumbel (G),
survival Gumbel (SG), Clayton (C), survival Clayton (SC), Normal (N) and the independence copula (I)\footnote{Obviously, for the independence copula no parameter needs to be
estimated.}. We then select the best fitting copula family from these 11 candidate models by the Bayesian information criterion (BIC). Given the
size and complexity of our model, as well as the difficulty to estimate parameters precisely on higher trees,
we decided to rely on the BIC to find more parsimonious model specifications and to minimize the estimation error. Additionally, we considered the restriction
of allowing potential time variation only on a limited (small) number of trees. We considered this restriction from 1 to 12 trees, but we report only a subset of
these models since results are identical or at least very similar for many of those cases. Specifically, it turned out that making this restriction
beyond the 6$^{th}$ tree is irrelevant, since there is no evidence of time variation on higher trees. A further possible restriction that may be made
is to truncate the vine beyond a certain tree, meaning that all conditional copulas are set to the independence copula.
In the current application we did not make this restriction because the independence copula is included in the set of admissible models and the automatic
selection by the BIC in practice leads to a truncation. Nevertheless, for large dimensional applications truncation of the vine should definitely be considered.

\begin{table}[ht]
  \centering
  \begin{tabular}{rrrrrrr}
  \hline
 &  M0& M1&M2& M3&M4 & M6\\
time-varying&   -&1   &1-2 &1-3 &1-4 &1-6\\
time-constant&1-27&2-27&3-27&4-27&5-27&7-27\\
    \hline
up to tree      1 &  -11491& -12177&-12177&-12177&-12177&-12177\\
 up to tree     2 &  -13657& -14313&-14346&-14346&-14346&-14346\\
up to tree      3 &  -14893& -15482&-15517&-15525&-15525&-15525\\
up to tree      4 &  -15653& -16207&-16239&-16265&-16265&-16265\\
up to tree      5 &  -16068& -16538&-16563&-16583&-16584&-16584\\
up to tree      6 &  -16326& -16774&-16787&-16796&-16802&-16804\\
up to tree      7 &  -16466& -16882&-16892&-16892&-16909&-16914\\
up to tree     20 &  -17358& -17621&-17599&-17580&-17585&-17594\\
total   &  -17366& -17626&-17604&-17587&-17593&-17603\\
     \hline
   \end{tabular}
   \caption{Partial BIC (up to trees 1-7 and 20) and total BIC values for Models with no (M0), first tree (M1),
   first and second (M2) , first three (M3), first four (M4)  and first six trees (M6) time-varying}
   \label{tab:daxbic}
 \end{table}

\begin{table}[ht]
  \centering
  \begin{tabular}{r|rr|rrrrrr|r}
  \hline
  \multicolumn{10}{l}{M0: all trees time-constant}\\
  Tree & \multicolumn{2}{c|}{time-varying} & \multicolumn{6}{c|}{time-constant } & \# par  \\
       & N & SG & I & N  & C  & G & SC & SG  &             \\
     \hline
      1       & - & -  & 0  & 7  & 0 & 4  & 0  & 17   & 28  \\
      2       & - & -  & 0  & 10 & 0 & 12 & 0  & 5    & 27  \\
      3       & - & -  & 2  & 14 & 0 & 7  & 0  & 3    & 24  \\
      4       & - & -  & 2  & 15 & 0 & 4  & 1  & 3    & 23 \\
      5       & - & -  & 4  & 15 & 0 & 4  & 0  & 1    & 20  \\
   \hline
  \multicolumn{10}{l}{M1: tree 1 time-varying, trees 2-27  time-constant}\\
    1 &  23 & 4  & 0  & 0  & 0 & 0 & 0  & 1  &  82  \\
    2 &  -  & -  & 0  & 4  & 0 & 8 & 0  & 15 &  27 \\
    3 &  -  & -  & 1  & 9  & 0 & 5 & 1  & 10 &  25  \\
    4 &  -  & -  & 2  & 13 & 0 & 7 & 0  & 3  &  23  \\
    5 &  -  & -  & 5  & 16 & 0 & 2 & 1  & 0  &  19  \\
    \hline
   \multicolumn{10}{l}{M2: trees 1-2 time-varying, trees 3-27  time-constant}\\
      1 &  23 & 4  & 0  & 0  & 0 & 0 & 0  & 1  &  82  \\
      2 & 1  & 2  & 0  & 4  & 0 & 8 & 0  & 12 & 33  \\
      3 & -  & -  & 1  & 8  & 0 & 5 & 1  & 11  & 25 \\
      4 & -  & -  & 2  & 14 & 0 & 6 & 0  & 3  & 23  \\
      5 & -  & -  & 5  & 15 & 0 & 2 & 1  & 1  & 19 \\
      \hline
   \multicolumn{10}{l}{M3: trees 1-3 time-varying, trees 4-27  time-constant} \\
      1 &  23 & 4  & 0  & 0  & 0 & 0 & 0  & 1  &  82  \\
      2 & 1  & 2  & 0  & 4  & 0 & 8 & 0  & 12 & 33  \\
      3 & 3  & 0  & 1  & 7  & 0 & 5 & 1  & 9  &  31  \\
      4 & -  & -  & 2  & 10 & 0 & 7 & 0  & 6  & 23  \\
      5 & -  & -  & 5  & 14 & 1 & 2 & 1  & 1  &  19  \\
       \hline
   \multicolumn{10}{l}{M4: trees 1-4 time-varying, trees 5-27  time-constant}\\
      1 &  23 & 4  & 0  & 0  & 0 & 0 & 0  & 1  &  82  \\
      2 & 1  & 2  & 0  & 4  & 0 & 8 & 0  & 12 & 33  \\
      3 & 3  & 0  & 1  & 7  & 0 & 5 & 1  & 9  &  31  \\
      4 & 1  & 0  & 2  & 9  & 0 & 7 & 0  & 6  &  25 \\
      5 & -  & -  & 5  & 14 & 1 & 2 & 1  & 1  &  19 \\
      \hline
   \multicolumn{10}{l}{M6: trees 1-6 time-varying, trees 7-27  time-constant} \\
      1 &  23 & 4  & 0  & 0  & 0 & 0 & 0  & 1  &  82  \\
      2 & 1  & 2  & 0  & 4  & 0 & 8 & 0  & 12 & 33  \\
      3 & 3  & 0  & 1  & 7  & 0 & 5 & 1  & 9  &  31  \\
      4 & 1  & 0  & 2  & 9  & 0 & 7 & 0  & 6  &  25 \\
      5 & 0  & 0  & 5  & 14 & 1 & 2 & 1  & 1  &  19  \\
      6 & 1  & 0  & 11 & 6  & 1 & 1 & 0  & 3  &  14  \\
      7 & -  & -  & 8  & 4  & 1 & 3 & 0  & 6  &  14  \\
   \end{tabular}
   \caption{Number of bivariate copula families used with time-varying and time-constant
   parameters and total number of parameters in trees 1-5 for models allowing for increasing number
   of trees with time-varying parameters}
   \label{tab:daxcopula}
 \end{table}

The results of our estimation presented in Tables \ref{tab:daxbic} and \ref{tab:daxcopula} report the estimation results when allowing time variation in
no, in trees 1 to 4 and in tree 6, respectively. For trees 5 and 7-28 no time-varying dependence was found. Table \ref{tab:daxcopula} reports the number of
time-varying copulas found on each tree and how often each copula family was selected. In addition, the number of parameters estimated in each tree is
given. In Table \ref{tab:daxbic} partial and total BIC values are reported. We note that time variation is very important, when modeling the dependence
in the first tree.  Here 27 out of 28 copulas are chosen to be time-varying. Furthermore, the vast majority of the time-varying copulas is selected to be
the Gaussian copula, which is in line with the findings \citet{HM10}. An explanation for this finding is given in \citet{MS11}, who show that the Gaussian
copula with random correlations has much larger dependence in the tails than the static Gaussian copula. This is in line with the stylized fact that
financial returns are characterized by tail dependence. Beyond the first tree, however, only very few pairs have time-varying dependence justifying the
restrictions. Among the copulas with constant parameters all families are chosen, but Gaussian, survival Gumbel and independence copulas are selected more often than the
other families. Especially on higher trees the independence copula dominates, indicating that a truncation after a certain level (say 15-20) would come
almost without any costs in terms of model fit. The overall fit of the models measured by the BIC for the total model turns out to be best when allowing
time variation only for on the first level. This can be explained by the strong penalty on additional parameters by the BIC and the better fit on the
higher trees.

\begin{figure}[ht!]

    \begin{center}
        \subfigure{%
            \label{fig:tv_tau1a}
            \includegraphics[width=0.48\textwidth]{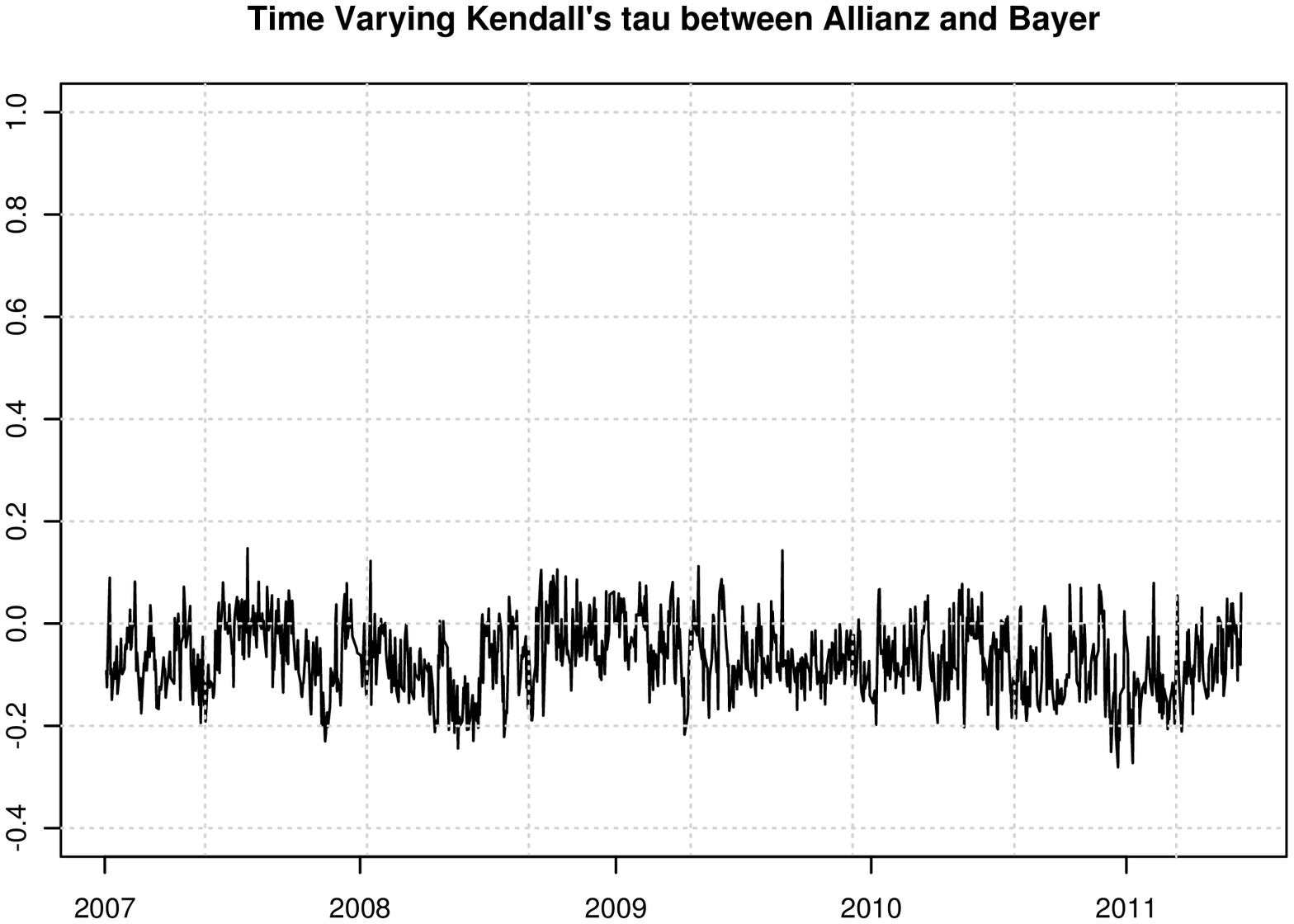}
        }%
        \subfigure{%
           \label{fig:tv_tau1b}
           \includegraphics[width=0.48\textwidth]{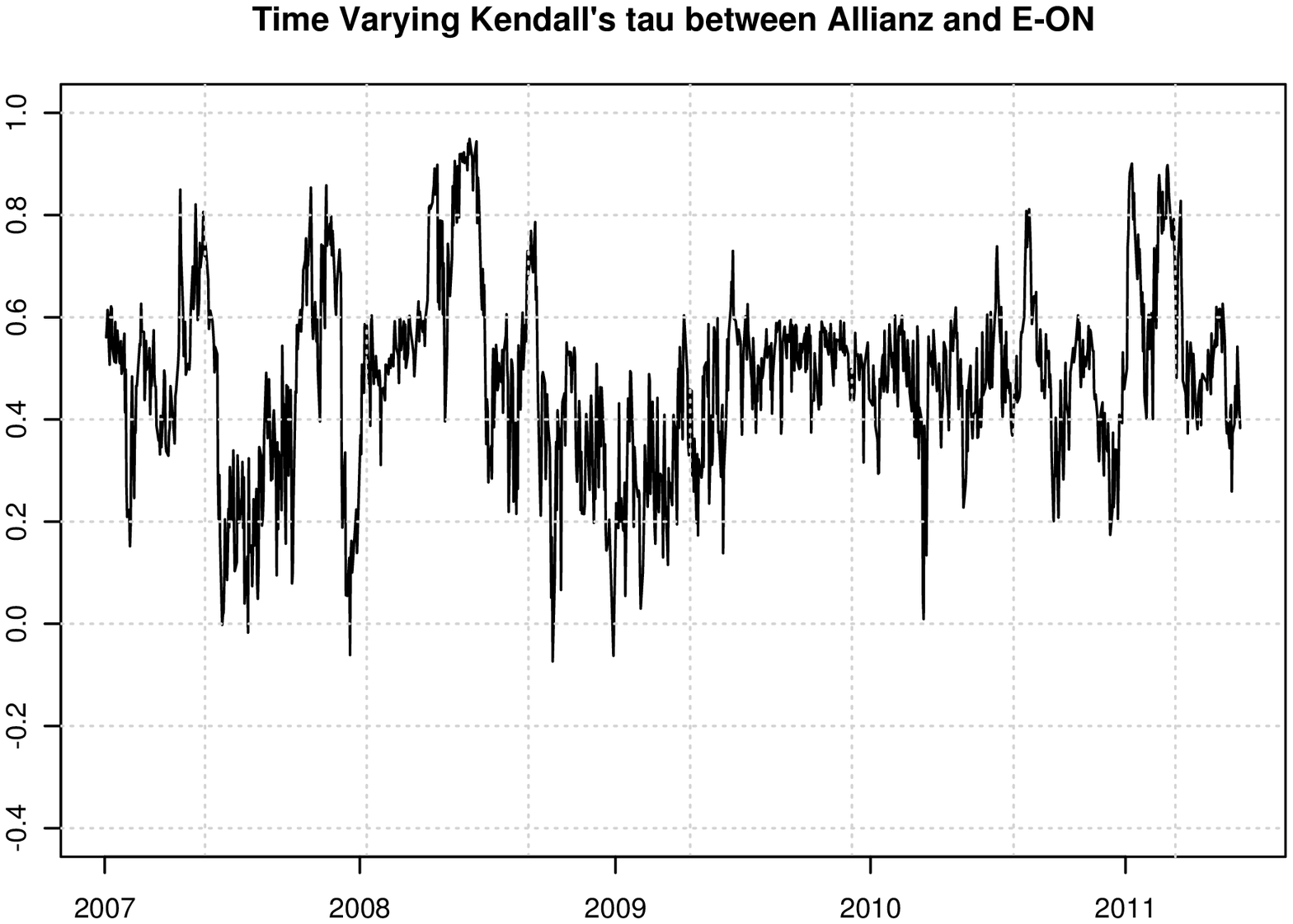}
        }\\ 
        \subfigure{%
            \label{fig:tv_tau1c}
            \includegraphics[width=0.48\textwidth]{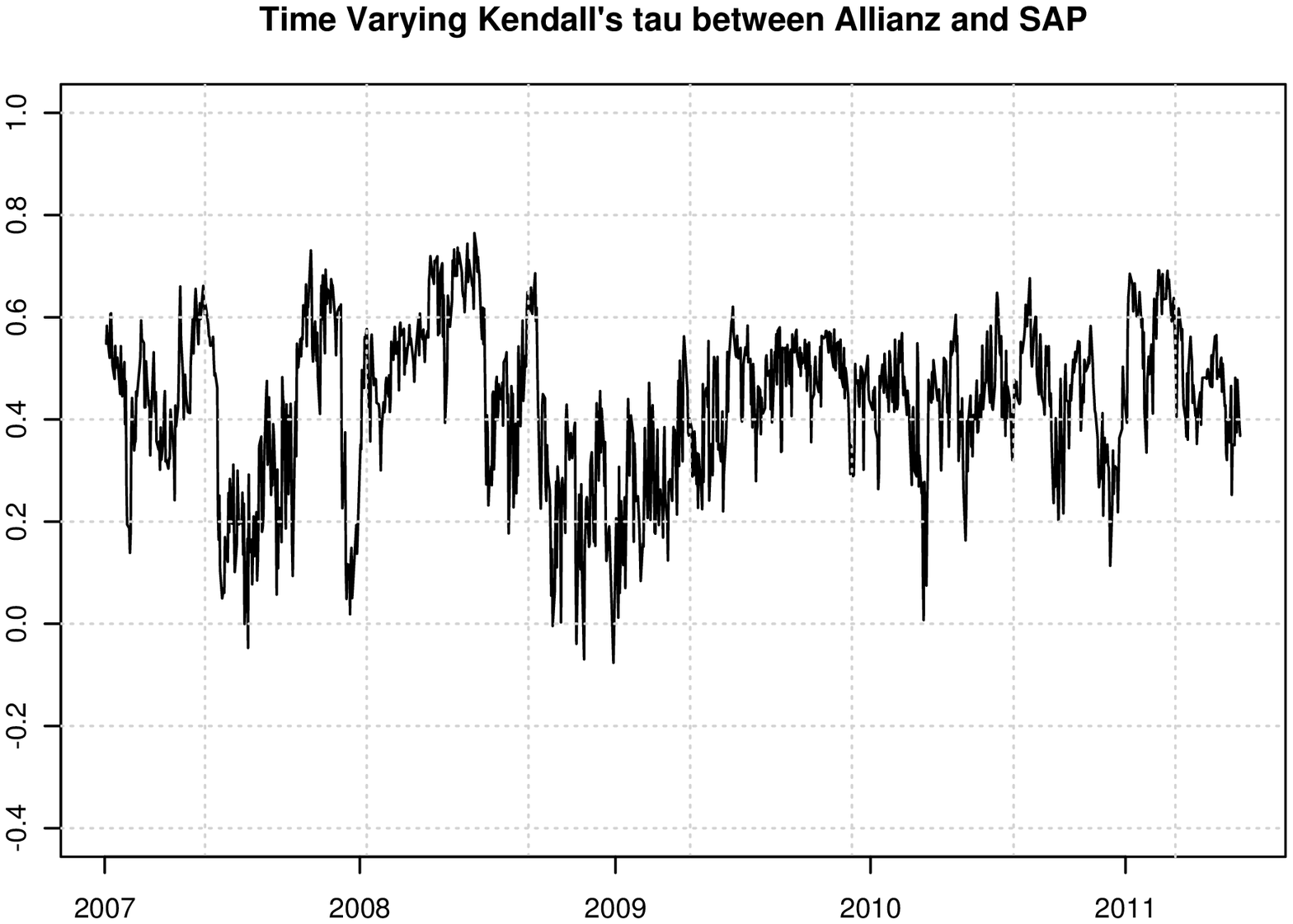}
        }%
        \subfigure{%
            \label{fig:tv_tau1d}
            \includegraphics[width=0.48\textwidth]{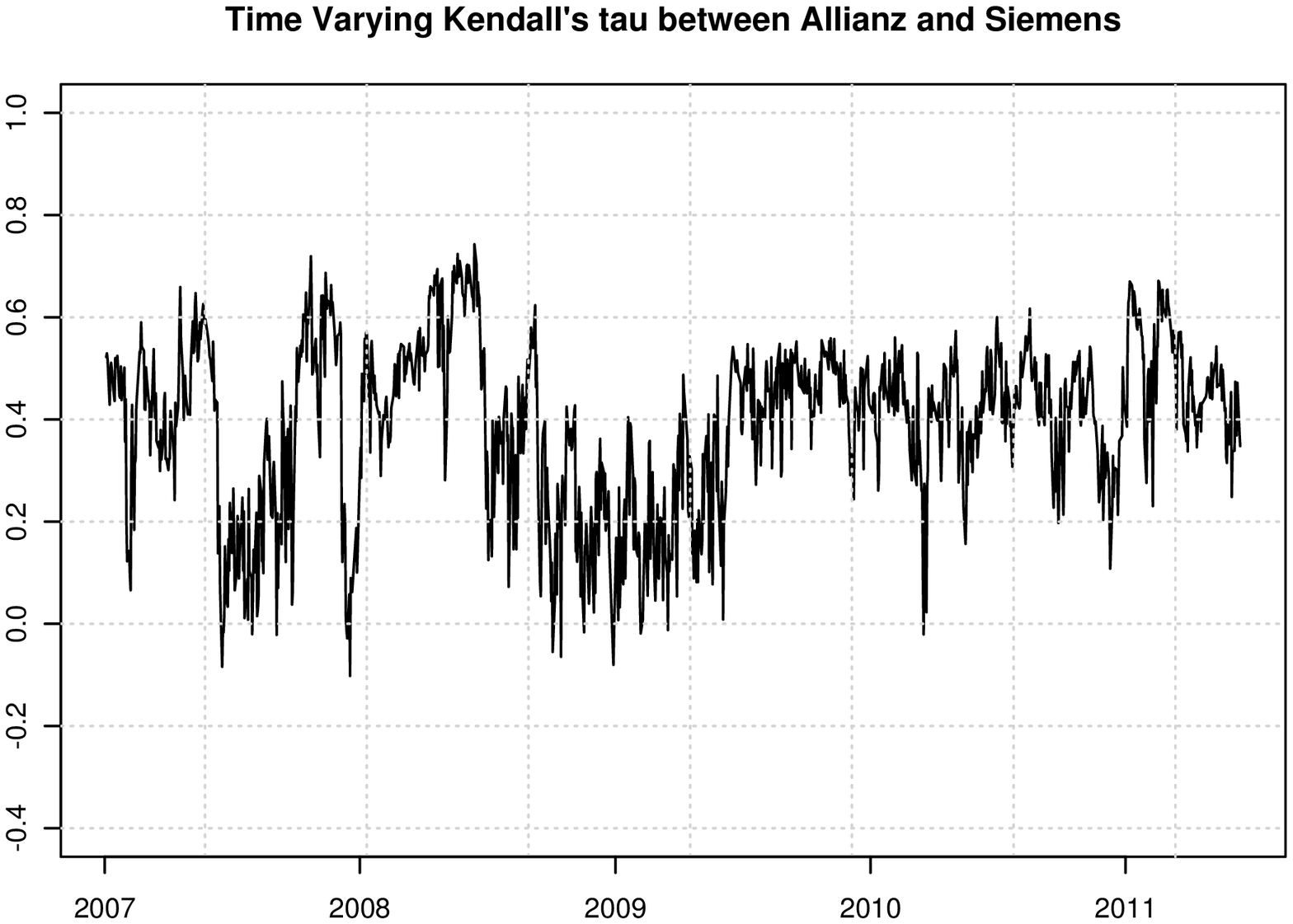}
        }\\%
        \subfigure{%
            \label{fig:tv_tau2a}
            \includegraphics[width=0.48\textwidth]{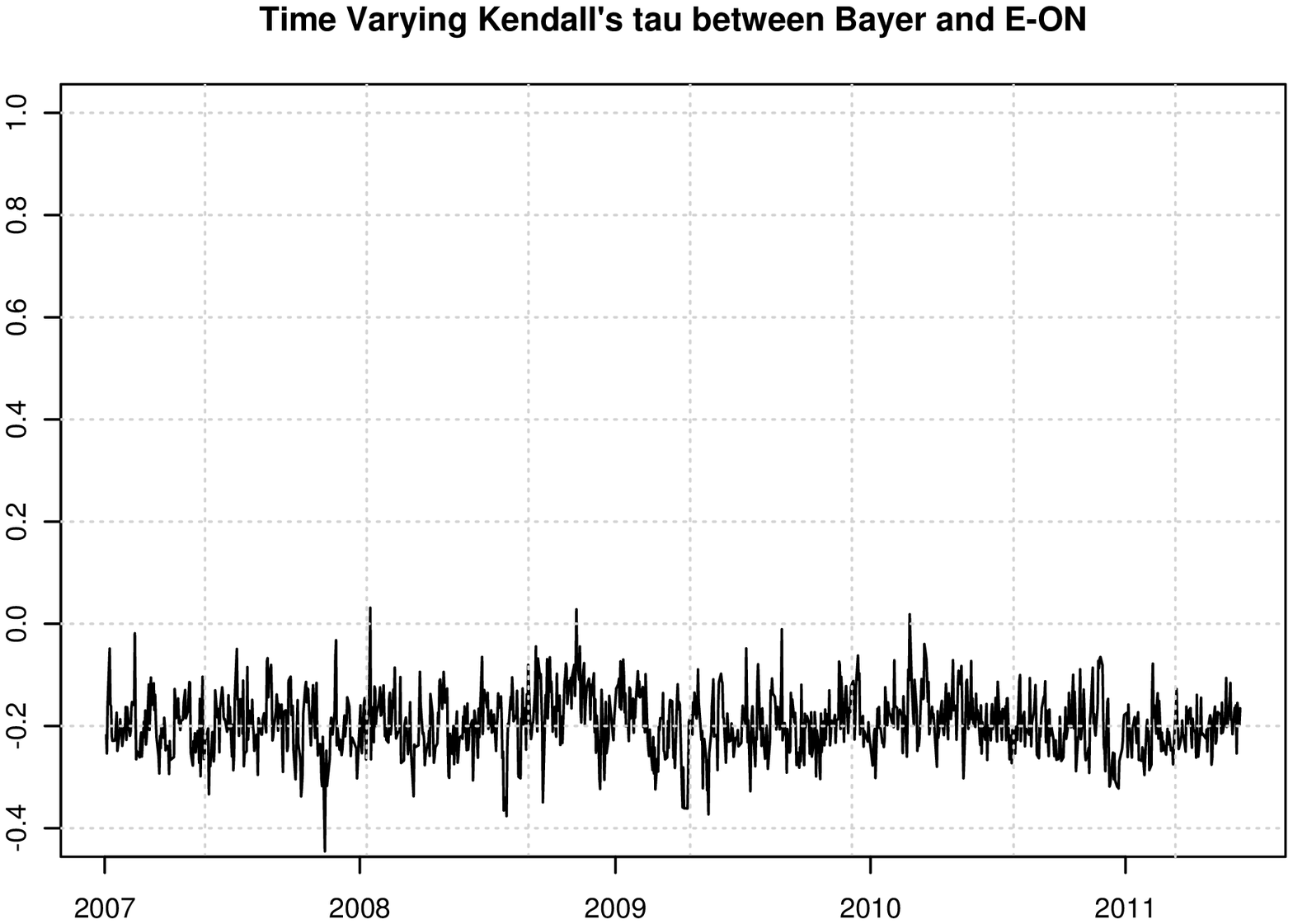}
        }%
        \subfigure{%
           \label{fig:tv_tau2b}
           \includegraphics[width=0.48\textwidth]{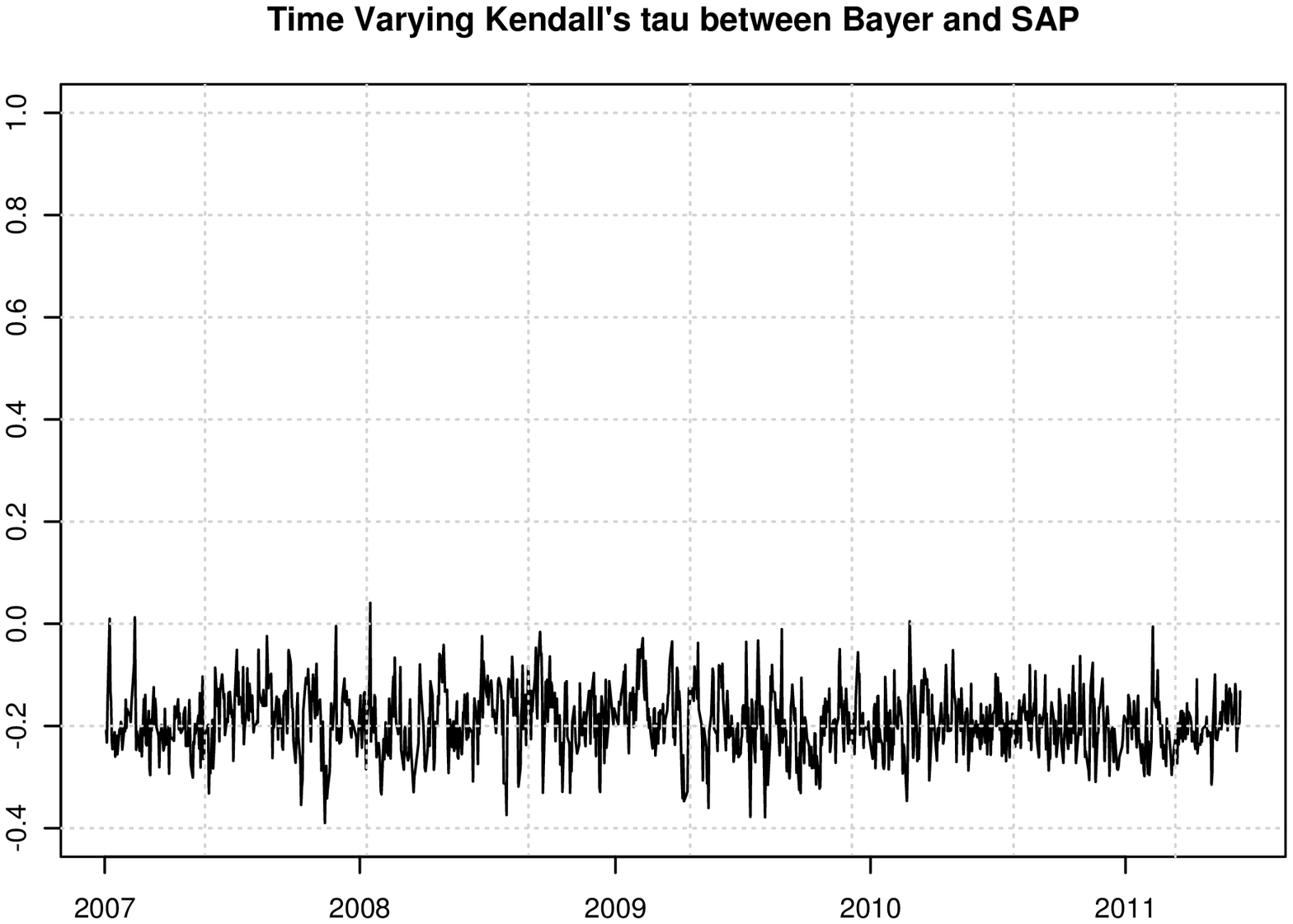}
        }\\ 

    \end{center}
    \caption{Smoothed estimates for time varying Kendall's $\tau$}
        \label{fig:tv_tau1}
\end{figure}

\begin{figure}[ht!]

    \begin{center}
        \subfigure{%
            \label{fig:tv_tau2c}
            \includegraphics[width=0.48\textwidth]{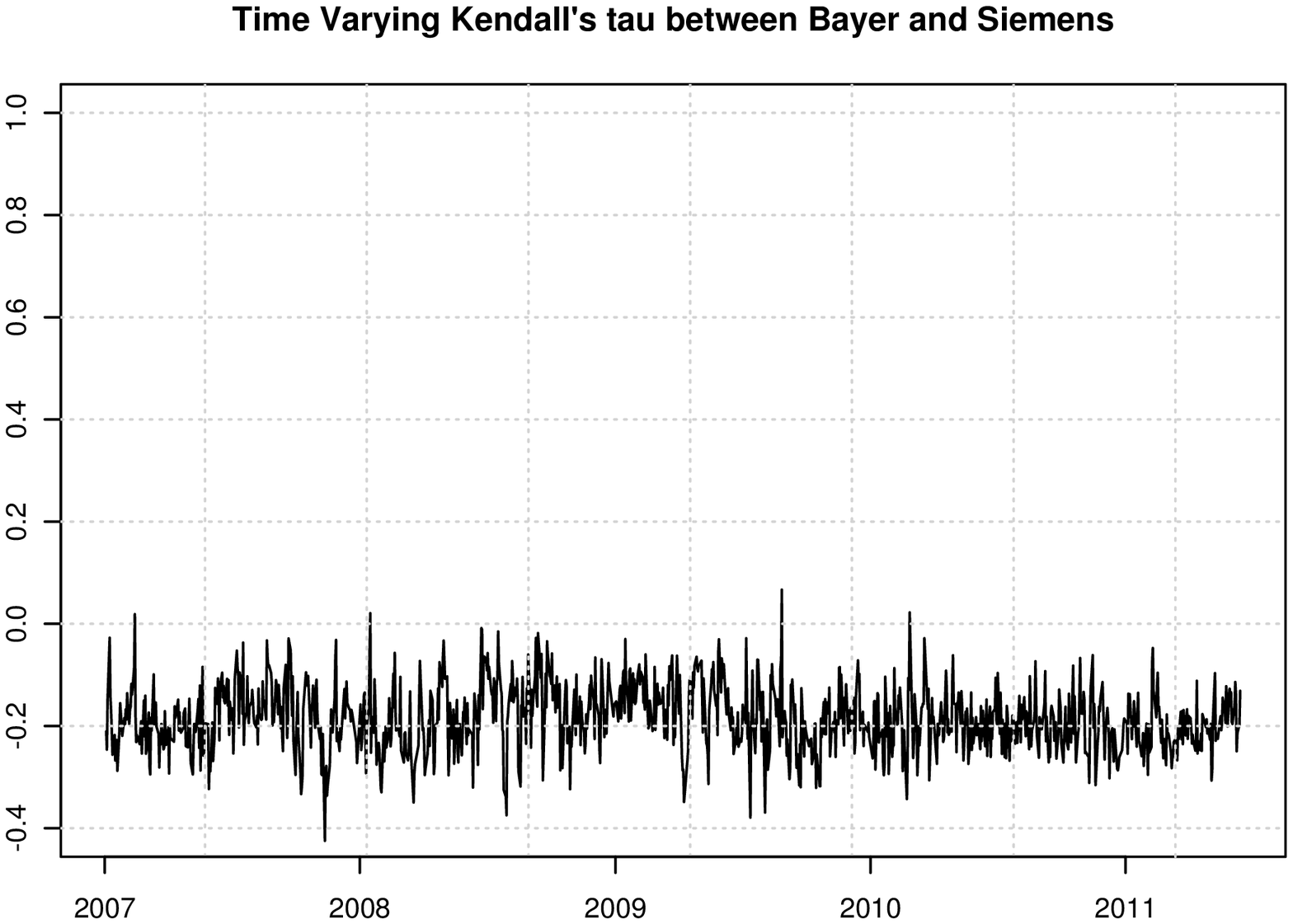}
        }%
        \subfigure{%
            \label{fig:tv_tau2d}
            \includegraphics[width=0.48\textwidth]{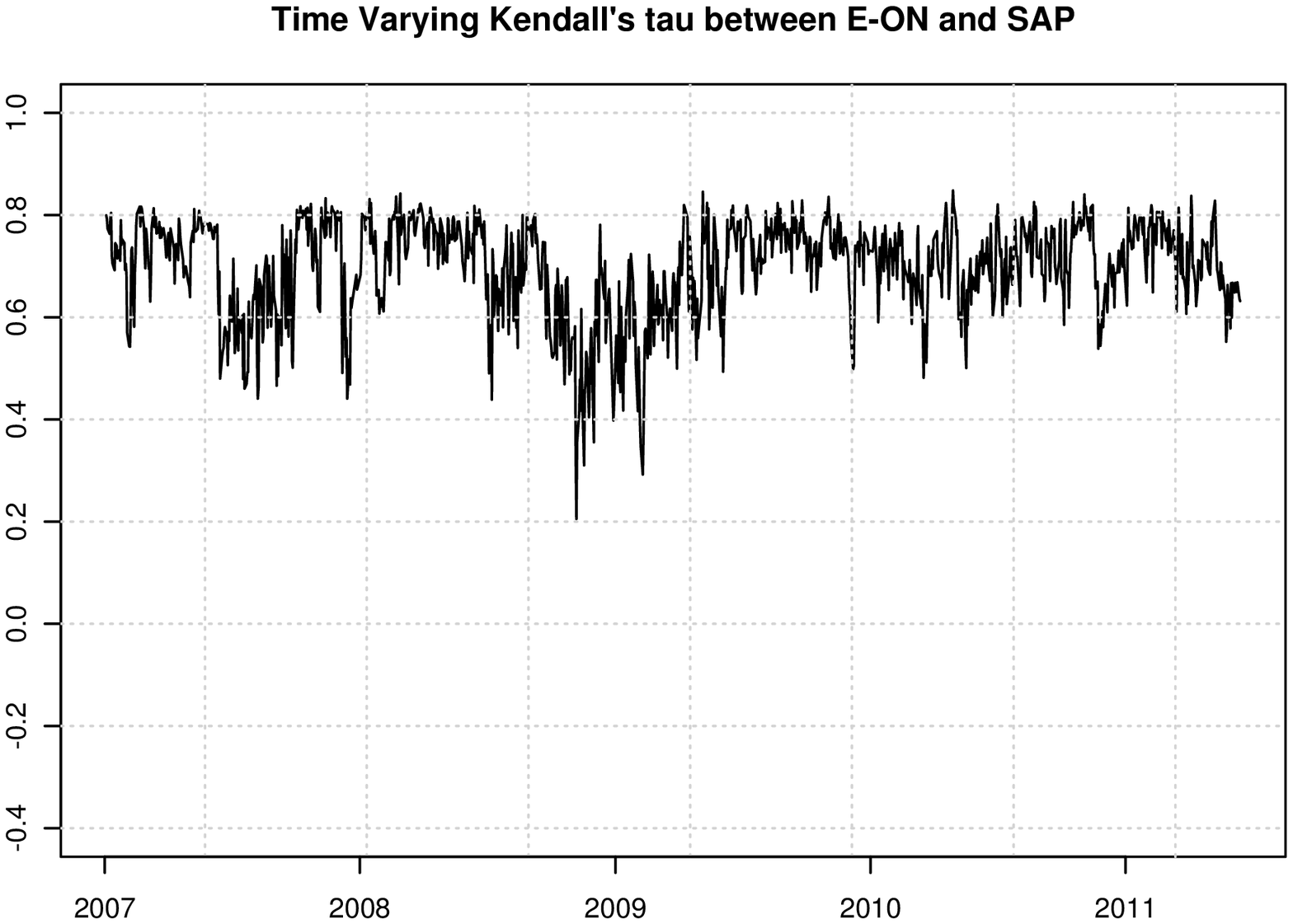}
        }\\%

        \subfigure{%
            \label{fig:tv_tau3a}
            \includegraphics[width=0.48\textwidth]{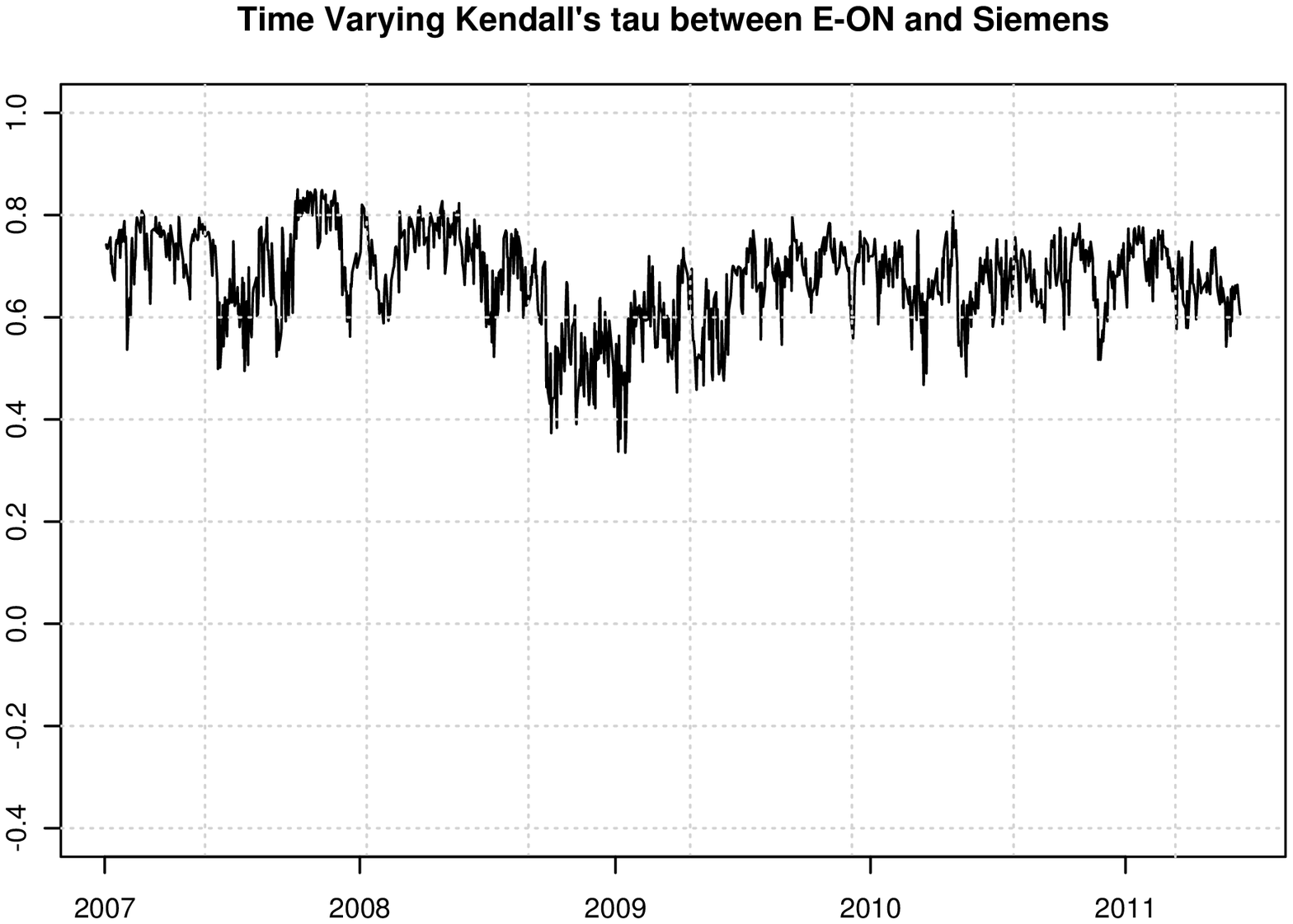}
        }%
        \subfigure{%
            \label{fig:tv_tau3b}
            \includegraphics[width=0.48\textwidth]{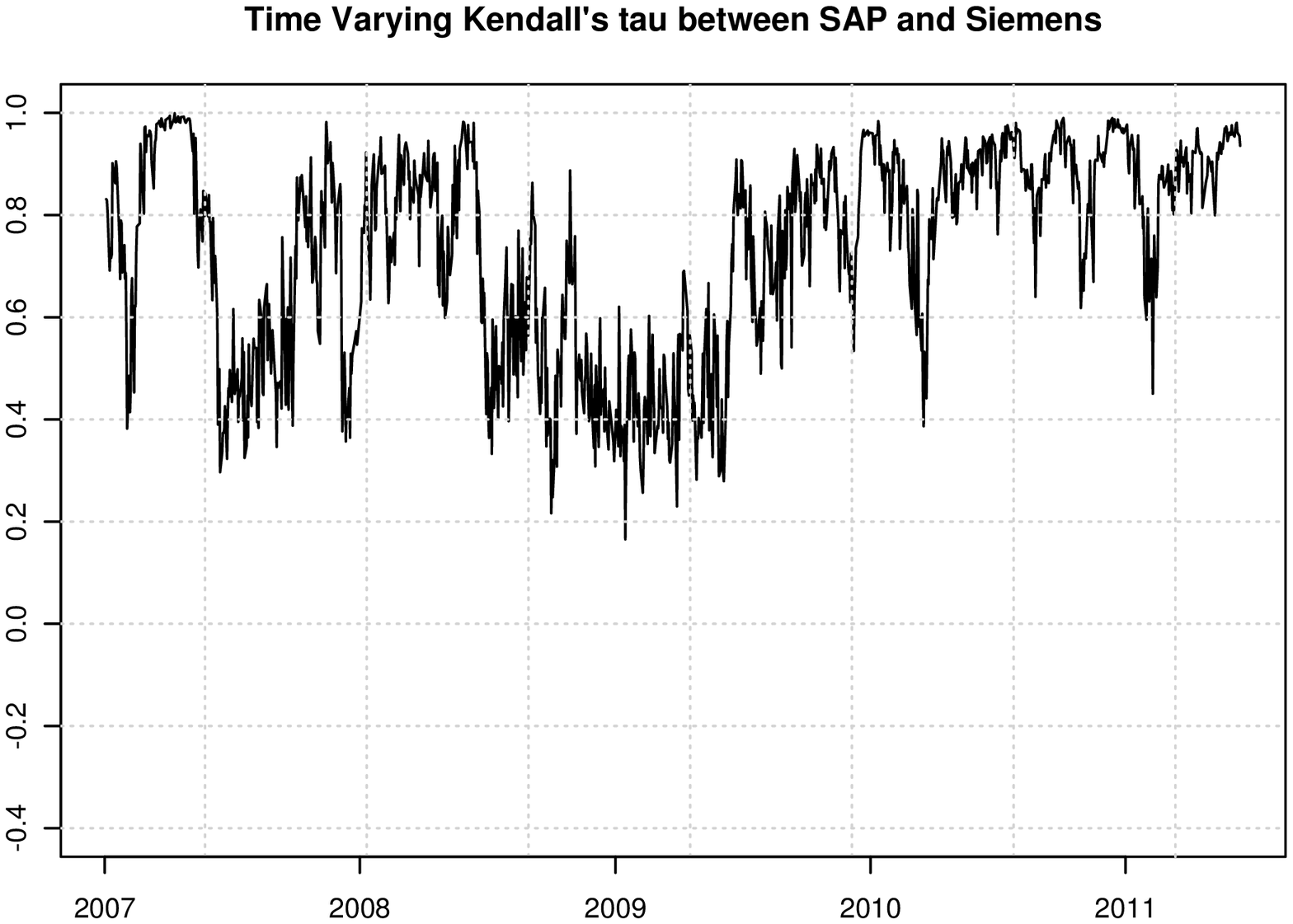}
        }%
    \end{center}
    \caption{Smoothed estimates for time varying Kendall's $\tau$}
        \label{fig:tv_tau2}
\end{figure}

Finally, we are interested in estimating the path of the pairwise dependence parameters.
Smoothed estimates for the time-varying Kendall's $\tau$ based on the model with time-variation only on the first tree are presented in Figures \ref{fig:tv_tau1} and \ref{fig:tv_tau2} for the pairwise dependence of the five companies that have the largest weight in the DAX index. The companies are (with their corresponding node in the D-vine in parenthesis) Allianz (13), Bayer (9), E.ON (12), SAP (27) and Siemens (19). This choice covers the interesting situations that we have companies that are neighboring in the vine, i.e. their dependence is modeled directly, that companies are close in the vine and that they lie quite far apart from each other. In the latter situation the implied pairwise time-varying dependence is computed conditional on the dependence between all variables in between. Note that in this case the dependence parameter at each point in time has to be computed using Monte Carlo simulation, for which we used 400 Monte Carlo replications. The dynamics of the dependence parameter are clearly visible for all situations and we have strong evidence of dependence changing over time. For example, the dependence parameters have decreased strongly in 2009 for most pairs. It also stands out that the dependence parameters involving the company Bayer have much less pronounced movements in dependence, which is mostly negative or close to independence.

\section{Conclusions and further research}

\label{Sec:Conclusion}

In light of the current financial crisis including the discussion of
understanding systemic risk the need to understand time-varying effects not
only within individual financial products but also among groups of financial variables
has been increasing. The developed D-Vine SCAR models are aiming to fill this
demand. From a statistician's point of view such a model is demanding. First a
very flexible multivariate dependency model is required such as the  class of
vine copulas and secondly an appropriate model for the time dependency of the
copula parameters has to found. For parameter interpretability we chose to
follow a parameter driven instead of an observation driven approach and thus
allowed for a stochastic autoregressive structure for the copula parameters to
model time-varying  dependence among a large group of variables.

While this approach leads to a relatively straightforward model formulation
the development of efficient estimation procedures is much more difficult.
Especially in high dimensions maximum likelihood is infeasible, since it would
require the maximization over integrals of size equal to the data length. In
the application the data length was 1125. These integrals occur since we need
to integrate over the latent variable process  to express the joint likelihood.
This problem already occurs when we consider bivariate SCAR models. One
solution to this is to use efficient importance sampling \citep{RZ07}. In
addition, the pair copula construction approach of \citet{ACFB09} for
multivariate copulas allows to express the likelihood in bivariate copula terms
in addition to a sequential formulation over the vine tree structure. This
makes it feasible to develop and implement efficient importance sampling for
the D-vine SCAR model. In a simulation study we validated our estimation
procedure and the application to joint dependency modeling of 29 stocks in the
DAX showed that time-varying dependence structures can be observed. One
interesting feature of this application is that nonnormal pair copulas with
constant parameters were replaced by normal pair copulas when time-varying
copula parameters are allowed. This was observed when the strength of the
dependence was moderate to large.

In this paper we followed some first approaches to model selection. We first
restricted to a known dependency structure given by a D-vine, but allowed the
copula family of each pair copula to be chosen among a prespecified class of
copula families in addition to the choice if a pair copula has time-varying
parameters or not. For this we used BIC, however more sophisticated criteria
might be necessary. We also restricted the use of time-varying pair copula
parameters to a prespecified number of top trees. Here the approach of
truncated vines as developed in \citet{brechmann-etal} might be a good starting
point to choose this number in a data driven manner. As already mentioned it is
feasible to extend the class of D-vine SCAR models to include R-vines as
copula models. More research is also needed to investigate the effects of time-varying copula models on economic entities such as portfolio returns and value
at risk. Finally the model uncertainty introduced by assuming the marginal
parameter estimates as true ones in the two-step approach has to assessed.
However the simulation results of \citet{Kim2007} might remain valid for a
copula model with time-varying parameters.

\clearpage
\begin{appendix}
\section{Stocks in the DAX and their ordering}
\begin{table}[h!]
\small{
\begin{center}
\begin{tabular}{llr}
Company          &      Ticker Symbol & node in D-vine\\
\hline
Adidas               & ADS  &25       \\
Allianz              & ALV   & 13      \\
BASF                 & BAS   & -      \\
Bayer                & BAYN  &9      \\
Beiersdorf           & BEI   &6      \\
BMW                  & BMW    & 18     \\
Commerzbank          & CBK    &15     \\
Daimler              & DAI    &17     \\
Deutsche Bank        & DBK    &14     \\
Deutsche Börse       & DB1   &24      \\
Lufthansa            & LHA    &16     \\
Deutsche Post        & DPW     & 22    \\
Deutsche Telekom     & DTE    &10     \\
E.ON                 & EOAN  & 12      \\
Fresenius            & FRE    &3     \\
Fresenius Medical Care &  FME  &2    \\
HeidelbergCement   &   HEI     &26    \\
Henkel             &   HEN3    &7    \\
Infineon   Technologies   &  IFX& 28   \\
K+S                &   SDF   &4      \\
Linde               &   LIN   &8     \\
MAN                 &   MAN    &21    \\
Merck               &   MRK    &1    \\
Metro               &   MEO    &5    \\
Munich Re           &   MUV2   &23    \\
RWE                 &   RWE     & 11   \\
SAP                 &   SAP     & 27   \\
Siemens             &   SIE     &19   \\
ThyssenKrupp        &   TKA    &20    \\
Volkswagen Group    &   VOW3    & 29   \\
\end{tabular}
\caption{DAX companies and their node position in the selected D-vine for the sample period from the 1$^{st}$ of January 2007 until the 14$^{th}$ of June 2011.
The company BASF(BAS) was dropped from the analysis, because data was not available for the complete sample period.}

\end{center}}
\end{table}

\end{appendix}

\clearpage

\bibliography{MVSCAR_bibliography}

\end{document}